\documentclass[11pt,preprint,graphicx]{aastex}
\usepackage{graphicx}

\def\etal{\it et al. \rm }

\begin{document} 

\title{Colors of Ellipticals from {\it GALEX} to {\it Spitzer}}

\author{James M. Schombert$^{A,B}$}
\affil{$^A$Department of Physics, University of Oregon, Eugene, OR USA 97403}
\affil{$^B$jschombe@uoregon.edu}

\begin{abstract}

\noindent Multi-color photometry is presented for a large sample of local ellipticals
selected by morphology and isolation.  The sample uses data from $GALEX$, SDSS, 2MASS
and {\it Spitzer} to cover the filters $NUV$, $ugri$, $JHK$ and 3.6$\mu$m.  Various
two-color diagrams, using the half-light aperture defined in the 2MASS $J$ filter,
are very coherent from color to color, meaning that galaxies defined to be red in one
color are always red in other colors.  Comparison to globular cluster colors
demonstrates that ellipticals are {\it not} composed of a single age, single
metallicity (e.g., [Fe/H]) stellar population, but require a multi-metallicity model
using a chemical enrichment scenario.  Such a model is sufficient to explain
two-color diagrams and the color-magnitude relations for all colors {\it using only
metallicity as a variable on a solely 12 Gyrs stellar population with no evidence of
stars younger than 10 Gyrs}.  The [Fe/H] values that match galaxy colors range from
$-$0.5 to +0.4, much higher (and older) than population characteristics deduced from
Lick/IDS line-strength system studies, indicating an inconsistency between galaxy
colors and line indices values for reasons unknown.  The $NUV$ colors have unusual
behavior signaling the rise and fall of the UV upturn with elliptical luminosity.
Models with BHB tracks can reproduce this behavior indicating the UV upturn is
strictly a metallicity effect.

\end{abstract}

\section{Introduction}

One of the most intriguing topics in extragalactic research is the stellar
populations in galaxies.  For the underlying stellar population of a galaxy's light
reveals the star formation and chemical history of galaxies, as well as representing
the dominant baryon component in most galaxies.  It is through the stars in galaxies
that we observe them at high redshift, and how we understand the large scale
structure of the Universe and the evolution of galaxies from the formation epoch to
the present.  Understanding galaxy kinematics, $M/L$'s, IMF, morphology and formation
scenarios all revolve around the properties and characteristics of a galaxy's stellar
population.

Ellipticals represent one of the most carefully studied type of galaxies with respect
to stellar populations.  They exhibit the simplest morphological and structure as
well as internal kinematics and, thus, are well-modeled as a single stellar
population rather than the kinematically distinct components found in disk galaxies.
They typically occupy the highest masses (i.e., luminosities) and, therefore, are the
clearest signposts at high redshift.  Studies of galaxy evolution often focus on
ellipticals owing to early indications that their spectrophotometric changes are the
simplest to model and to trace reliability through cosmic time.

The study of stellar populations in ellipticals has historically taken three
different routes.  The first is the use of optical and near-IR colors to interpret
the integrated light of the underlying stellar population.  The discovery of the
color-magnitude relation (CMR, Sandage \& Visvanathan 1978), the separation of
morphology types by color (Tojeiro \etal 2013) and different galaxy components by
color (e.g., bulge versus disk, Head \etal 2014) were early explorations into the
stellar populations and the meaning of color with respect to the star formation
history of galaxies (Tinsley 1978).  Technological improvements in the 1980's led to
the obvious extension of multi-color work through a higher inspection of the
spectroenergy distribution (SED) with study of various spectral indices related to
different types of stars found in a stellar population with a range of stellar
masses.  This type of investigation reached a peak with the development of the
Lick/IDS line-strength system (Worthey \etal 1994; Trager \etal 2000) where a set of
specific spectral features were should to correlate with the two primary
characteristics of a stellar population, its age and mean metallicity.  Guided by SED
models of the Lick/IDS line-strength system (see Graves \& Schiavon 2008), these
spectral indices became the observable of choice to study nearby and distant galaxy
stellar populations.  Lastly, with the launch of HST, space imaging provides the best
study of a stellar population by direct examination of their color-magnitude
diagrams, although this is still limited to the nearby Universe.  For a majority of
stellar population studies, the Lick/IDS line-strength system is the method of
choice.

The properties of optical colors of ellipticals, particularly the various
color-magnitude relations, have often been taken as evidence in favor of the
monolithic scenario for galaxy formation, the production of the entire stellar
population of a galaxy in one single burst at redshifts greater than 5. The lack of
redshift evolution of the slope and scatter in optical CMRs (e.g., van Dokkum \etal
2000), and observed  passive color evolution (Rakos \& Schombert 1995), is consistent
with a high formation redshift ($z > 2$).  However, the predicted star formation
histories (SFHs) of ellipticals in the hierarchical merger paradigm (Kauffmann \&
Charlot 1998; Khochfar \& Burkert 2003) are much more complicated.  For example, the
predicted SFHs of ellipticals in the merger scenario are expected to be
quasi-monolithic, with an overwhelming majority of the stellar mass forming before a
redshift of 1 (Kaviraj \etal 2005) and expected to be difficult to discriminate using
optical colors given the well-known age-metallicity degeneracy problem (Worthey
1994).

The use of colors for investigating stellar populations in galaxies is needed even in
the spectroscopic era.  For example, higher signal-to-noise is acquired for faint,
distant galaxies using colors.  Large areal surveys can be obtained by wide-field
cameras using well chosen filter sets.  In the 1990's, Rakos \& Schombert pioneered
the use of narrow band filters selected to cover age/metallicity features around the
4000\AA\ break as an fast and efficient system to study cluster galaxies using
imaging (see Rakos \& Schombert 1995).  The results from those studies confirmed
a passive evolution for the stellar populations in cluster ellipticals, but was in
sharp disagreement with the results from spectroscopic surveys that found much
younger ages and lower metallicities for the same objects (Trager \etal 2000, Graves
\etal 2009, Conroy \etal 2014).  Larger SDSS samples (Gallazzi \etal 2005, Graves
\etal 2010) presented a wider and more diverse range of ages and metallicities, and
more sophisticated analysis techniques (see Johansson \etal 2012; Conroy \etal 2014;
Worthey \etal 2014) reinforced the trend of age and metallicity with luminosity and
mass.

The disagreement between color results and spectroscopic results was outlined in
Schombert \& Rakos (2009), an analysis of the CMR in the cluster environment and the
expected colors based on the measured ages and metallicities with the Lick/IDS
line-strength system.  Despite the clear disparity between the expected colors from
the young, metal-poor stellar populations deduced from Lick/IDS indices, the system
is still very much in use (McDermid \etal 2015) and the resulting age and metallicity
estimates are the core to most theoretical scenarios for galaxy formation and
evolution (see review by Naab 2013).  Resolving the conflict between colors and line
indices is a crucial stellar population problem.

In addition, colors from the extreme ends of the UV to near-IR wavelengths
provides particularly useful information to various astrophysical problems.  For
example, the near and far-IR colors are salient to studies of the old component in
stellar populations and the baryonic mass to light ratio (Schombert \& McGaugh 2014)
The near and far-UV colors investigate the so-called "UV upturn" problem (Bertola
\etal 1982, Brown \etal 1997), the unusually high UV fluxes in ellipticals
presumingly dominated by old stellar populations without current star formation.
The past determination of UV colors has presented mixed results with positive 
(Donas \etal 2007) and negative (Jeong \etal 2009) CMR slopes, and the prediction
that only bright ellipticals display this behavior (Yi \etal 2005).

This paper presents a comprehensive analysis of elliptical colors using archival data
from the near-UV ({\it GALEX NUV} to the far-IR ({\it Spitzer} 3.6$\mu$m).  With a
near-IR selected sample from the Revised Shapley-Ames and Uppsala Galaxy catalogs, we
will explore the behavior of galaxy colors over a wavelength range provided by {\it
GALEX}, SDSS, 2MASS and {\it Spitzer}.  Comparison with previous color studies will
examine the differences and various photometric relationships.  The colors presented
herein will anchor the zeropoint of elliptical colors for use by high redshift
studies.   The colors will also provide a window in the stellar populations of
ellipticals by comparison with simple and multi-metallicity population models.  The
wide wavelength coverage offers an avenue to break the age-metallicity degeneracy
that plagues optical colors (Worthey 1994).  

\section{Sample}

The data for this study was based on the sample defined by Schombert \& Smith (2012),
a purely morphological sample of ellipticals selected from the taken from either the
Revised Shapley-Ames (RSA, a catalog selected by luminosity) and Uppsala Galaxy
Catalog (UGC, an angular limit catalog).  The sample was restricted to large
angular-sized ($D > 2$ arcmins) galaxies and must have been imaged by the 2MASS
project.  In addition, the sample had to satisfy a criteria of isolation from
foreground or background objects (i.e., there are no nearby bright stars or companion
galaxies that would distort the surface brightness isophotes).

The resulting $JHK$ surface photometry was presented in Schombert \& Smith (2012),
and the final sample consisted of 436 ellipticals.  That sample was then
cross-correlated with the {\it Galaxy Evolution Explorer} ({\it GALEX}, Martin \etal
2005), SDSS and {\it Spitzer} image libraries for existing data from 226.7nm ({\it
GALEX} $NUV$) to 3.6$\mu$m (channel 1, {\it Spitzer}).  Using automated scripts to
browse the various mission websites resulting in 2,925 image files from the four
missions.  A breakdown of the images is shown in Table 1.  Overall there were 436
ellipticals in the 2MASS sample, of which 149 had {\it GALEX} data, 263 had matching
SDSS images and 149 with archived {\it Spitzer} 3.6$\mu$m images.  Given varying sky
completeness of each mission, the number of possible cross match filter colors is
shown in Table 1.

Each object in the total sample was also inspected for evidence of emission lines
(excluding AGN features), dust or other signatures of recent star formation.  The
idea here was to find a sample of ellipticals that was as similar in terms of
morphology and star formation history as possible.  While some low level signature of
AGN activity was acceptable, their effects on the galaxy's colors had to be restrict
to inside the various mission images PSF.  In addition, for galaxies with {\it GALEX}
images, a color selection was made base on the criteria for recent star formation
from Schawinski \etal (2007) that $NUV-g < 5.5$, and this color cutoff was mapping
into $g-r$ and $J-K$.

All the images were ellipse subtracted to look for asymmetric features that might be
a signature of recent mergers or dust lanes.  While the usual selection of boxy-like
and disk-like residuals were observed, there were no obvious linear features.   In
addition, color subtracted frames were examined for an evidence of dust lanes, none
were detected in the UV and optical images.

We conclude that the final sample is as simple and pure a sample of ellipticals as
can be found in the local Universe (versus most spectroscopic studies which present a
mix of early-type galaxies, including S0's).  All the objects are between 10 and 200
Mpc in distance (use the Benchmark values for $H_o$, $\Omega_M$ and
$\Omega_{\Lambda}$).  Their total luminosities ranged from $-$26 to $-$19 in $J$,
enclosing the range of ellipticals that are called "bright" or "normal" to the giant
ellipticals.  No true dwarf ellipticals are in the sample, nor are any true brightest
cluster ellipticals (i.e., cD ellipticals, Schombert 1986) in the sample.

\subsection{Data Reduction}

Data reduction of the flattened, calibrated images from each mission was performed
with the galaxy photometry package ARCHANGEL (Schombert 2011).  These routines, most
written in Python, have their origin back to disk galaxy photometry from the late
1980's and blend in with the GASP package from that era (Cawson 1987).  The package
has four core algorithms that 1) aggressively clean and mask images, 2) fit
elliptical isophotes, 3) repair masked regions then perform elliptical aperture
photometry and 4) determine aperture colors and asymptotic magnitudes from curves of
growth and determine accurate errors based on image characteristics, such as the
quality of the sky value.

The photometric analysis of galaxies branches into four areas; 1) isophotal analysis
(the shape of the isophotes), 2) surface brightness determination and fitting (2D
images reduced to 1D luminosity profiles), 3) aperture luminosities (typically using
masked and repaired images and elliptical apertures) and 4) asymptotic or total
magnitudes (using curves of growth guided by surface brightness data for the halos,
see Schombert 2011).  Ellipticals are the simplest galaxies to reduce from 2D images
to 1D luminosity profiles since, to first order, they have uniformly elliptical
shaped isophotes (Jedrzejewski 1987).  Where many ellipticals display disky or boxy
isophotal shapes (Kormendy \& Bender 1996), this deviation is at the few percent
level and has a negligible effect on the surface brightness profile, aperture
luminosities or colors values.

Surface brightness determination and fitting consumes a large fraction of the
processing time for ellipticals.  Accurate surface brightness profiles require
detailed masking to remove foreground stars, nearby fainter galaxies and image
artifacts.  While cleaning an elliptical galaxy's image is simplified by the lack of
HII regions, dust lanes or other irregular features, the final accuracy of the
profile will be highly dependent on the quality of the data image, in particular the
flatness of the image and knowledge the true sky value.  The smooth elliptical shape
to early-type galaxies results in very low dispersions in intensity around each
isophote that, in turn, makes the removal of stellar and small background galaxies a
relatively simple and automated task.

Following the prescription outlined in Schombert \& Smith (2012), we processed all
the mission images in the same manner.  Despite the differing plate scales (i.e.,
arcsecs per pixel), orientations on the sky and flux calibrations, the {\it GALEX},
SDSS, 2MASS and {\it Spitzer} missions all provide well flattened final data
products, usually free of any obvious artifacts.  Very little image preparation was
required, other than confirming that the targets in the images were, in fact, the
correct galaxy (galaxy misidentification in the archive servers was not uncommon).
This was accomplished by comparison with the PSS-II J images at STScI/MAST and crude
luminosity estimates compared to the RC3 magnitudes (de Vaucouleurs \etal 1991).

Isophotal fitting on each image begins with a quick visual inspection of the field
for manual suppression of artifacts and marking the center of the galaxy.  Then, an
iterative ellipse fitting routine begins outside the core region, moving outward
fitting the best least-squares ellipse to each radius until the isophote intensity
drops below 1\% of sky.  The routine then returns to the core to finish the inner
pixels in a like manner.  During the ellipse fitting, pixels greater than (or less
than) 3$\sigma$ from the mean intensity are masked and removed using a 50\% growth
radius.  The resulting fits are output as mean intensity, dispersion around the
ellipse, major axis, eccentricity, position angle (and errors) plus the first four
intensity moments.  Conversion to surface brightness profiles uses the generalized
radius, the square root of the major times minor axis ($\sqrt{ab}$).  All spatial
parameters will be quoted using the generalized radii.

Resulting elliptical isophotes are calibrated (intensity and pixel size) using the
standard pipeline calibrations provided by the missions, then processed into surface
brightness profiles.  Various fitting functions have been applied to elliptical
surface brightness profiles over the years.   A full discussion of their various
strengths and weaknesses can be found in Schombert (2013).  In brief, the S\'{e}rsic
$r^{1/n}$ provides the best fits over the full range of surface brightness, but
suffers from coupling between its shape and characteristic radii parameters that
minimize its usefulness.  Templates are stronger match to the shape of elliptical
profiles (Schombert 2015), but are only parametrized by luminosity and do not provide
any structural metrics.  In the end, we found that empirically determined parameters,
such as half-light radius ($r_h$) and mean surface brightness ($<\mu>$), are the most
strongly correlated parameters with luminosity or stellar mass.

As this study is primarily concerned with colors, the determination of luminosity in
a consistent and accurate manner from the datasets is of the highest priority.
Aperture luminosities are calculated using the ellipses determined by the isophote
routines.  Care was taken to make sure that the same eccentricities and position
angles were used across the various mission images.  None of the sample galaxies
display any variation in eccentricity or position angle at the 2\% level from the
near-UV to the far-IR.  Thus, aperture values were, in effective, determined using
fixed radii in kpcs.  

The total luminosity of a galaxy is a much more problematic value to determine.  The
procedure used here is outlined in Schombert (2011), where elliptical apertures are
determined by a partial pixel routine from the masked images where the masked regions
have been filled by the local mean isophote.  While it seems obvious that masked
regions would reduce the calculated flux inside an aperture, in fact, this effect is
rarely more than 5 to 10\% the total flux of an elliptical.  However, this effect is
unlike Poisson noise in that it always works to reduce the measured luminosity.  The
cleaned images are then integrated as a function of radius to produce curves of
growth.  An added feature is that as the outer aperture are integrated, their fluxes
are corrected by the mean isophotal intensity as given by either the raw surface
brightness profile or the S\'{e}rsic $r^{1/n}$ fit to the profile.  This reduces the
noise in the outer apertures and often produces a smoother convergence to a stable
total flux.

This produces three total luminosity values; one based on the raw pixel values in the
apertures, a second where the mean surface brightness is used to determine the flux
in the outer apertures (typically outside the 80\% level) and a third that uses the
profile fits to integrate the outer apertures.  For 92\% of the galaxies in this
sample, the difference in three difference total magnitudes was less than 0.5\%.  In
the remaining cases, a majority were galaxies near the edge of frames where the
surface brightness profile was required to complete the missing flux.  The remaining
handful simply had unusual luminosity profiles, whether from instrumental errors or
astrophysical causes (e.g., mergers) could not be determined and they were eliminated
from the sample.

For a majority of the missions, the error quoted at the archives for the total and
aperture magnitudes severely underestimates the actual error found in this study (see
\S2.2, 2.3, 2.4 and 2.5).  This is due to the fact that their error calculations
focus, primarily, on the Poisson noise that is proportional to device sensitivity
and exposure time.  However, for large extended sources ($D > 2$ arcmins), the
primary source of noise is uncertainty in the sky value and the variation of sky
across the image.  Sky for this study was determined by two separate algorithms.  The
first is the manual section of between 10 and 20 sky boxes (typically 20 by 20 pixels
in size) in regions surrounding the target (outside its halo), but separate from
other galaxies or bright star halos.  The dispersion on the mean from averages in
each sky box provides the best value for the uncertainty on the sky value (Schombert
2013).  Total errors quoted in this paper are then calculated as 3$\sigma$ from the
mean sky value applied to the sum of the pixels in each aperture.  As a check to the
correct sky value, the elliptical isophotes are fixed in shape and tabulated beyond
the galaxy radius to the edge of the frames.  These outer ellipse intensity values
should converge on the sky value determined from the sky boxes.  In the few cases
where the two values disagree, the ellipse sky value was used, but the error
estimates continued to use the dispersion between the sky boxes.

Colors, the main focus of this study, were determined by two different methods.  The
first is the traditional fashion of directly comparing fluxes between two identically
sized apertures.  For regularly shaped objects like ellipticals, this is a simple
matter of picking a metric radius and interpolating the elliptical apertures in the
filters of interest.  As stated above, the ellipse shape were similar between the
various images from the UV to the IR.  The greatest uncertainty is associated with
the centroid position from frame to frame.  Again, the symmetric shape of ellipticals
makes centroid errors small.  And this effect will only be important for the colors
in very small apertures while attempting to estimate a core color.  To avoid this
problem, core colors will be determined by other methods.

Determination of color by comparing total luminosities in different filters
introduces too much error from sky noise to be a reliable measure.  Rather than
extrapolating to a total color value, most of the colors quoted in this study will be
some aperture color at a fractional luminosity, e.g. half-light radius.  As color
gradients are well-known in ellipticals (Kim \& Im 2013), some analysis of the
changes with different fractional luminosity radii will be presented in \S3.1.  For
consistence, the half-light radius for the sample will be determined in the
2MASS $J$ frame, the only filter with data for all the galaxies in the sample.  The
total luminosity (and, thus, half-light luminosity) are also fixed to the $J$ filter.

Full surface brightness analysis is performed on all the filter images.   Thus, a
second determination of color is possible by examining color as measured by the
difference in the surface brightness profiles directly.  While any particular color
surface brightness isophote contains much more error than the comparable aperture
color at the same radius, at large radii this method can be more informative as the
number of pixels used is large compared to the sky error.  Of course, color gradients
are the primary use for multi-filter surface brightness profiles.  And, again, with
the regularity of shape for ellipticals, the run of color with radius is a direct
measure of the projected 2D distribution of stellar populations.  The best measure of
the core color of an elliptical is the interpolation of the color surface brightness
profiles to zero radius.  Through out this paper we will refer to aperture colors by
their fractional radius as a subscript (e.g. $g-r_{0.5}$) and extrapolated colors
from surface brightness profiles with a letter designation (e.g. $g-r_o$ for a core
color or $g-r_h$ for a half-light color).

Apparent uncorrected luminosities are designated by lowercase letters (e.g.,
$m_{NUV}$ for raw {\it GALEX} $NUV$ luminosities).  Absolute luminosities, radii and
colors are corrected for Galactic extinction following the prescription of  Cardelli,
Clayton \& Mathis (1989) and $E(B-V)$ values from NED.  Magnitude units varied
between the missions; {\it GALEX}, SDSS and {\it Spitzer} use the AB/Vega system
($NUV,u,g,r,i,3.6$) and 2MASS used the Johnson system ($J,H,K$).  We have noted in
our discussions when the various units are used.  Distances used are the 3K CMB
distances using the Benchmark model values for the standard cosmological constants
(in particular, $H_o = 75$ km/sec/Mpc).  As none of the galaxies in this sample have
redshifts greater than 0.04, no k-corrections were applied to the data.  The final
data products are too extensive to be listed in this publication.  Instead, the
author maintains all the data, reduction scripts and log files at his website
(http://abyss.uoregon.edu/$\sim$js).  We follow the philosophy of presenting all the
reduction techniques as user enabled scripts, rather than a detailed description of
the various steps leading from raw images to final luminosities and colors.

\subsection{{\it GALEX}}

The {\it Galaxy Evolution Explorer} ({\it GALEX}, Martin \etal 2005) was a NASA Small
Explorer launched in April 2003.  It used a 50 cm aperture telescope with a dichroic
beam splitter that enabled simultaneous observations in the FUV ($\lambda_{eff}$ =
1539\AA, FWHM = 269\AA) and the $NUV$ ($\lambda_{eff}$ = 2316\AA, FWHM = 616\AA). The
field of view of the camera is circular with a 1.2 degree field of view.  The FWHM
varies within the field of view, but is, in general, 5 arcsecs for the central
region.  Full details of the data processing and calibration are given in Martin
\etal (2005), Morrissey \etal (2005) and Morrisey \etal (2007).

Processed images, obtained from MAST, are flattened, sky-subtracted, calibrated
frames with plate scales of 1.5 arcsecs per pixel.  A majority of the frames were
from dedicated programs (such as the Nearby Galaxy Survey and the Medium Imaging
program, see Gil de Paz \etal 2007), and many were the co-adds from several
overlapping programs.  Exposure times varied from 1,500 to 70,000 secs, 90\% of the
sample had exposure times greater than 3,000 secs.  Of the 436 galaxies in the 2MASS
defined sample, 149 had {\it GALEX} $NUV$ imaging.  A smaller percentage had
additional FUV imaging; however, that subsample was deemed too small in size and too
high in luminosity errors to be of use for this study.

\begin{figure}[!ht]
\centering
\includegraphics[scale=0.75,angle=0]{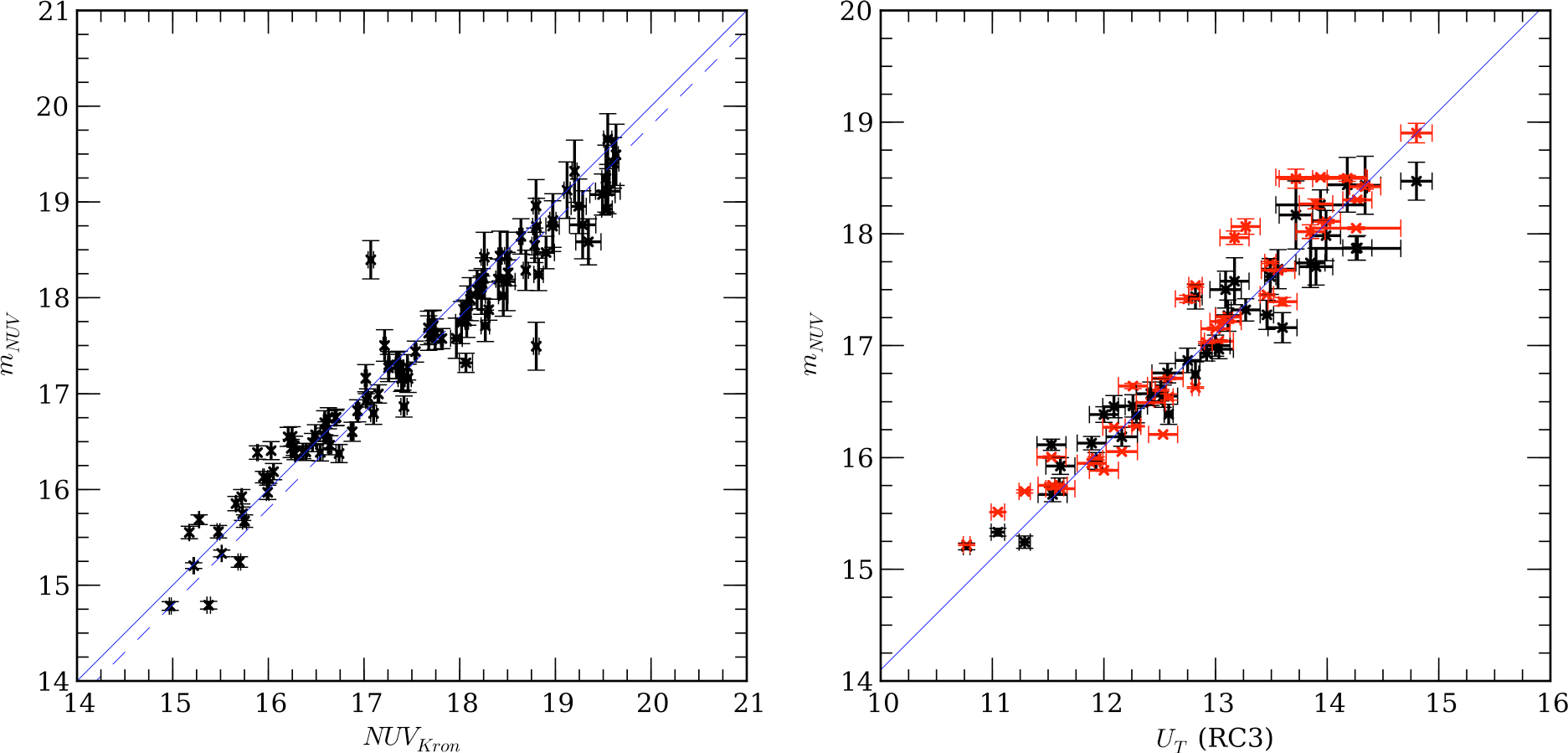}
\caption{\small Comparison of this study's $NUV$ luminosities with the {\it GALEX} projects
Kron magnitudes and Johnson $U$ from the RC3.  The left panel displays the comparison
for 103 ellipticals in common with the current {\it GALEX} database and our co-added $NUV$
frames.  The blue line indicates unity, the dashed line represents a measured
luminosity of 80\% total, as predicted by the Kron system.  The dispersion is 0.31
between the Kron magnitudes and our own measurements, in agreement with the mean
errors.  The right panel displays both datasets (black for this study, red for {\it GALEX}
Kron magnitudes) versus $U_T$ from the RC3.  The blue line represents a color difference
of $NUV-U=4.2$.
}
\label{arch_galex}
\end{figure}

Of the 149 analyzed ellipticals, 103 were also analyzed by the {\it GALEX} pipeline.
Photometry from the {\it GALEX} project tabulates the luminosity of galaxies in terms
of Kron magnitudes (Kron 1980).  The Kron magnitude system uses a luminosity weighted
radius to define a aperture for measuring a total flux.  Depending on the shape of
the surface brightness profile, this flux will be somewhere between 85 and 95\% the
total luminosity of the galaxy.  For ellipticals, a value of 90\% was been estimated
(Bertin \& Arnouts 1996).  The resulting comparison between {\it GALEX} Kron
magnitudes and this study's total luminosity ($m_{NUV}$ based on asymptotic fits) is
shown in the left panel of Figure \ref{arch_galex}, where both magnitudes are
calibrated to the AB system.  The blue line is a line of unity, the dashed line
displays the correlation if the Kron magnitudes represent 85\% the total flux.  The
agreement is fair, there is an apparent effect that Kron magnitudes capture more
total flux at brighter apparent magnitudes.  But, this effect is only 0.05 mags at
$m_{NUV}=19$ $NUV$.  The dispersion around the 85\% line is 0.31 mags, that is
similar to the average error in the $m_{NUV}$ values.  The average error quoted for
the {\it GALEX} Kron magnitudes is 0.03, which seems unrealistic.

The left panel of Figure \ref{arch_galex} displays comparison between $m_{NUV}$
(black symbols) and {\it GALEX} Kron magnitudes (red symbols) versus the $U_T$
magnitudes from the RC3 (de Vaucouleurs \etal 1991).  The blue line represents a
conversion of $U_T-NUV=4.2$.  For the 43 galaxies in common, the scatter for the
$m_{NUV}$ magnitudes is slightly smaller than the {\it GALEX} Kron magnitudes, but
both sets are within the errors of this study and the RC3.  The errors quoted by the
{\it GALEX} project are clearly too small for the observed scatter, probably owing to
the calculation by the {\it GALEX} pipeline using only Poisson noise rather than
correctly adopting the sky error into their error budget.  The average $NUV-U$ color
is 4.01$\pm$0.24, with a small color-magnitude effect visible for the small RC3
sample.

\subsection{SDSS}

The Sloan Digital Sky Survey (SDSS-III, Eisenstein \etal 2011) is an imaging and
spectroscopic survey of the northern sky using the 2.5m telescope at Apache Point
Observatory.  Imaging is obtained using a drift scanning technique in a five filter
system with centers at 3543\AA\ ($u$), 4770\AA\ ($g$), 6231\AA\ ($r$), 7625\AA\ ($i$)
and 9134\AA\ ($z$), closely following the original Gunn filter system (Oke \& Gunn
1983).  The current release for SDSS-III is DR12 (Alam \etal 2015) where footprint
images can be download from the SDSS website.  Each frame is 2048x1489 pixels with a
plate scale of 0.396 arcsecs per pixel.  Exposure time is 53.9 secs per pixel across
all filters.  Flux calibration is provided in nanomaggies and is included in each
image header.

Since SDSS is not an all sky survey, only 263 ellipticals from the 2MASS sample
overlap with the SDSS sample.  In addition, 106 SDSS ellipticals are imaged by {\it
GALEX} and 89 by {\it Spitzer}.  The final image frames are flattened and corrected
for atmospheric extinction.  Our data reduction pipeline worked on the DR12 frames in
the same manner as if they were raw telescope images.  Each frame was manually
inspected and cleaned of large objects.  Our final products were total luminosities,
aperture colors and a full surface brightness profile in each filter.

\begin{figure}[!ht] 
\centering 
\includegraphics[scale=0.75,angle=0]{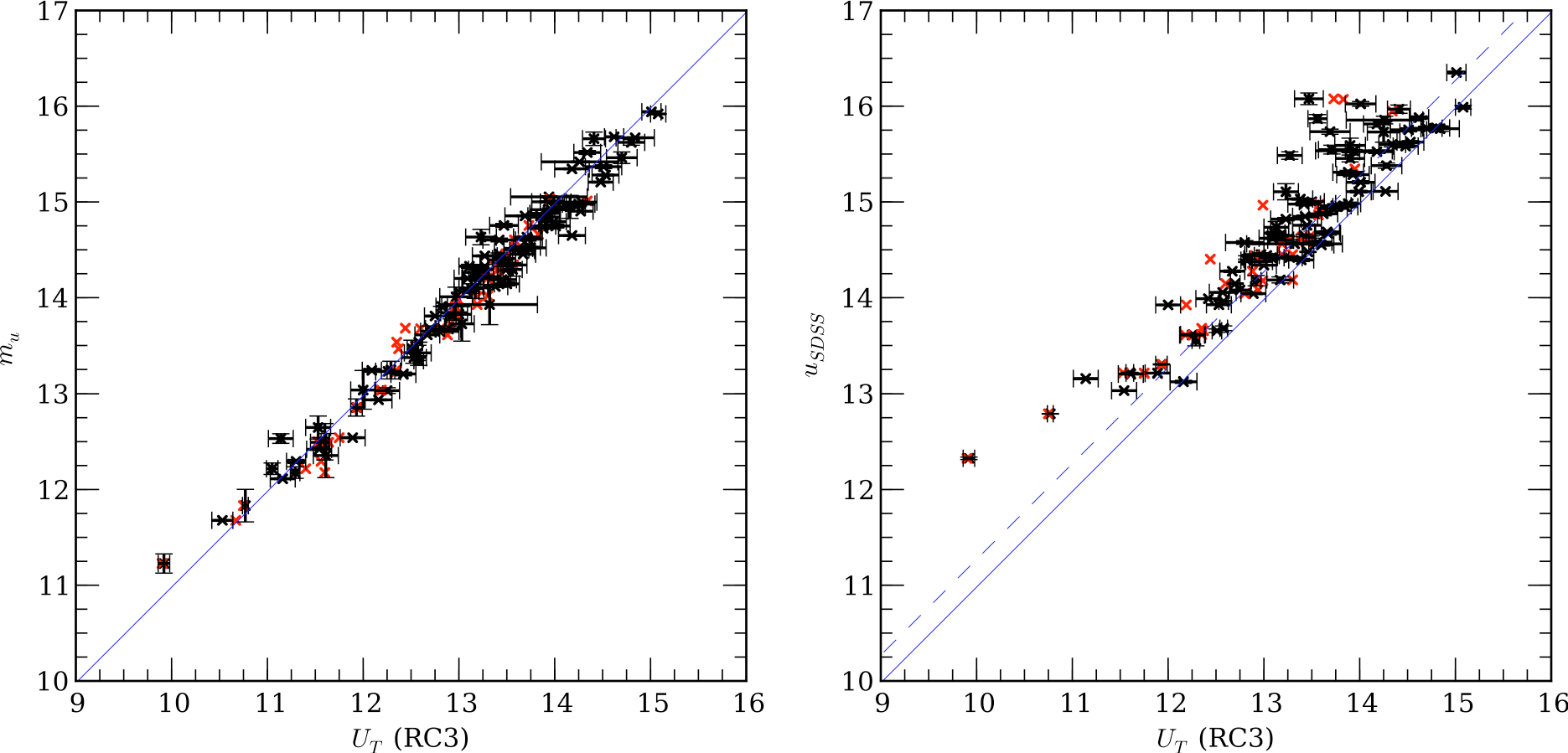}
\caption{\small Comparison of this study's $u$ luminosities versus the Johnson $U$
magnitudes and the SDSS Petrosian $u$ magnitudes.  The left panel display $m_u$ from
out study (uncorrected for Galactic extinction) versus Johnson $U$ from the RC3 (black
symbols) and Poulain \& Nieto (1994, hereafter PN94, red symbols).  The blue line
represents the standard Johnson $U$ to SDSS $u$ for the typical colors of an
elliptical.  The dispersion around the conversion line is 0.40 mags, consistent with
the quoted errors.  The right panel displays the DR12 Petrosian $u$ magnitudes
($u_{SDSS}$) versus RC3 and PN94.  The dashed line represents the expected relation
if the Petrosian magnitudes represent 80\% the total flux. The dispersion is 0.80
mags.
}
\label{petro_UT}
\end{figure}

\begin{figure}[!ht]
\centering
\includegraphics[scale=0.75,angle=0]{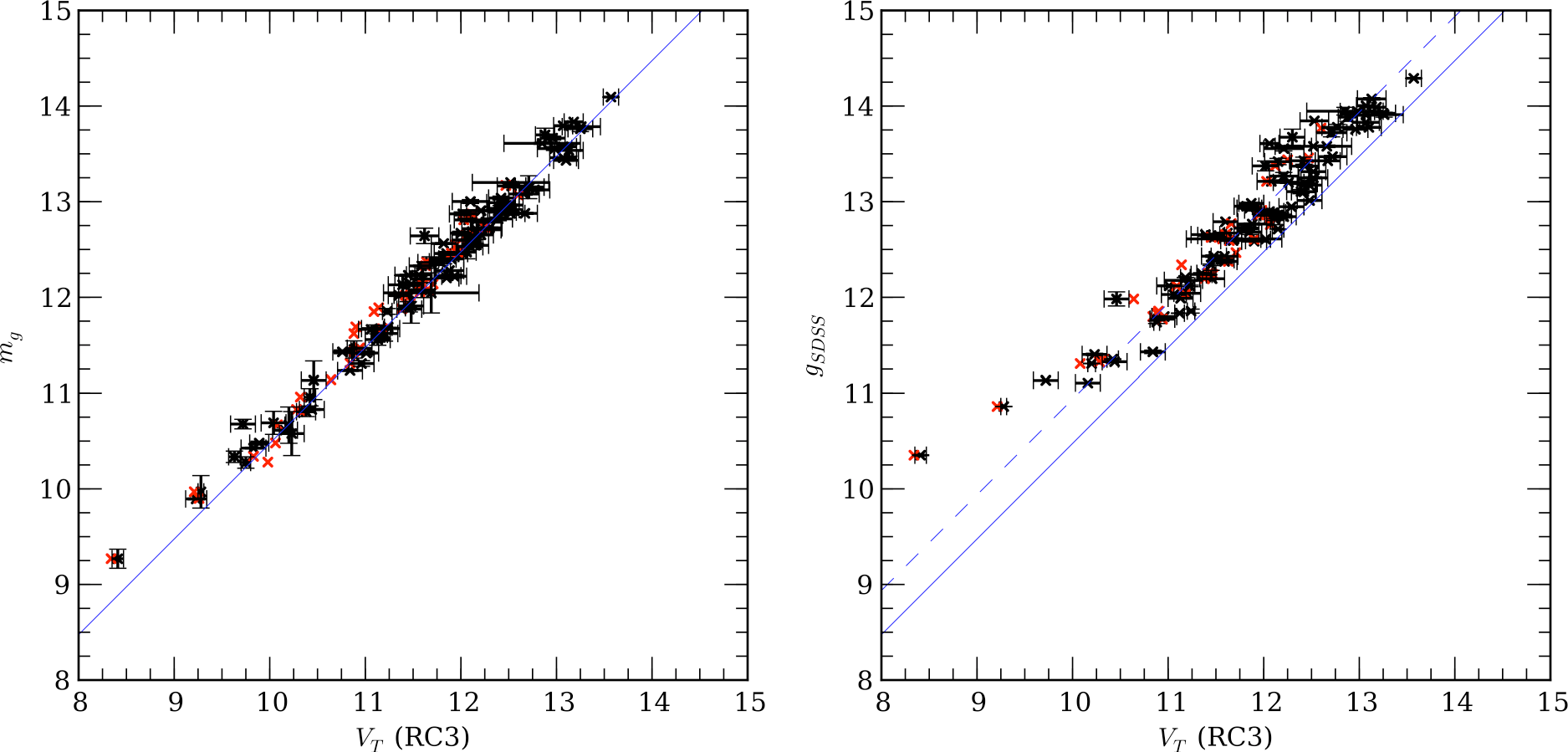}
\caption{\small 
The same as Figure \ref{petro_UT} for $g$ and Johnson $V$ from the RC3.  Symbols
and lines are the same as Figure \ref{petro_UT}, using $g$ to $V$ conversion factors.  
The dispersion is 0.15 mags for our luminosities, 0.28 mags for SDSS Petroisan
magnitudes.  Again, The dashed line represents the expected relation if the Petrosian
magnitudes represent 80\% the total flux.
}
\label{petro_VT}
\end{figure}

\begin{figure}[!ht] 
\centering 
\includegraphics[scale=0.75,angle=0]{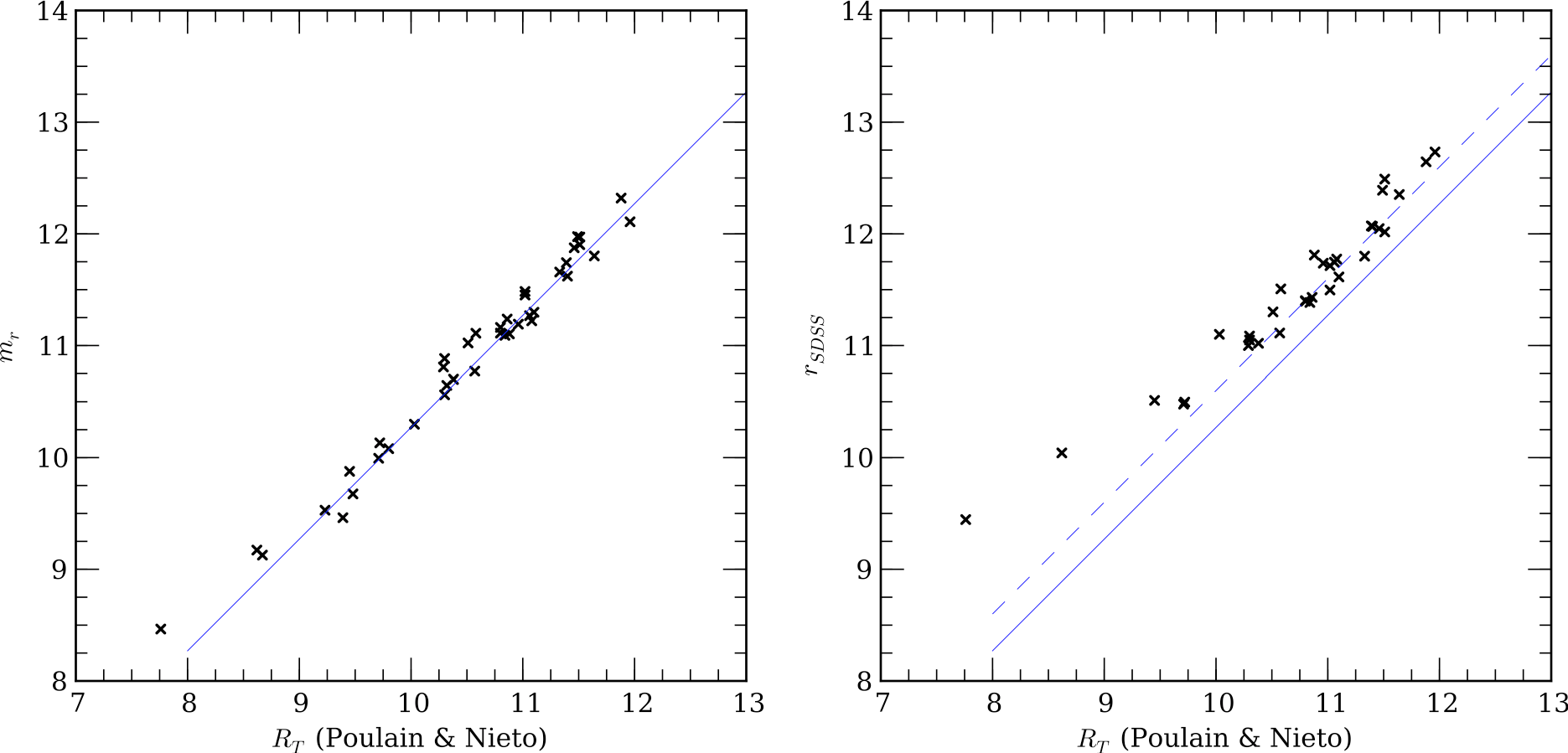}
\caption{\small The same as Figure \ref{petro_UT} for $r$ and Johnson $R$.  Symbols
and lines are the same as Figure \ref{petro_UT}, using $r$ to $R$ conversion factors.
The RC3 lacks Johnson $R$ measurements, so only limited data from Poulain \&
Nieto is displayed.  The dispersion is 0.14 mags for our luminosities, 0.18 mags for
SDSS Petroisan magnitudes.
}
\label{petro_RT}
\end{figure}

\begin{figure}[!ht]
\centering
\includegraphics[scale=0.75,angle=0]{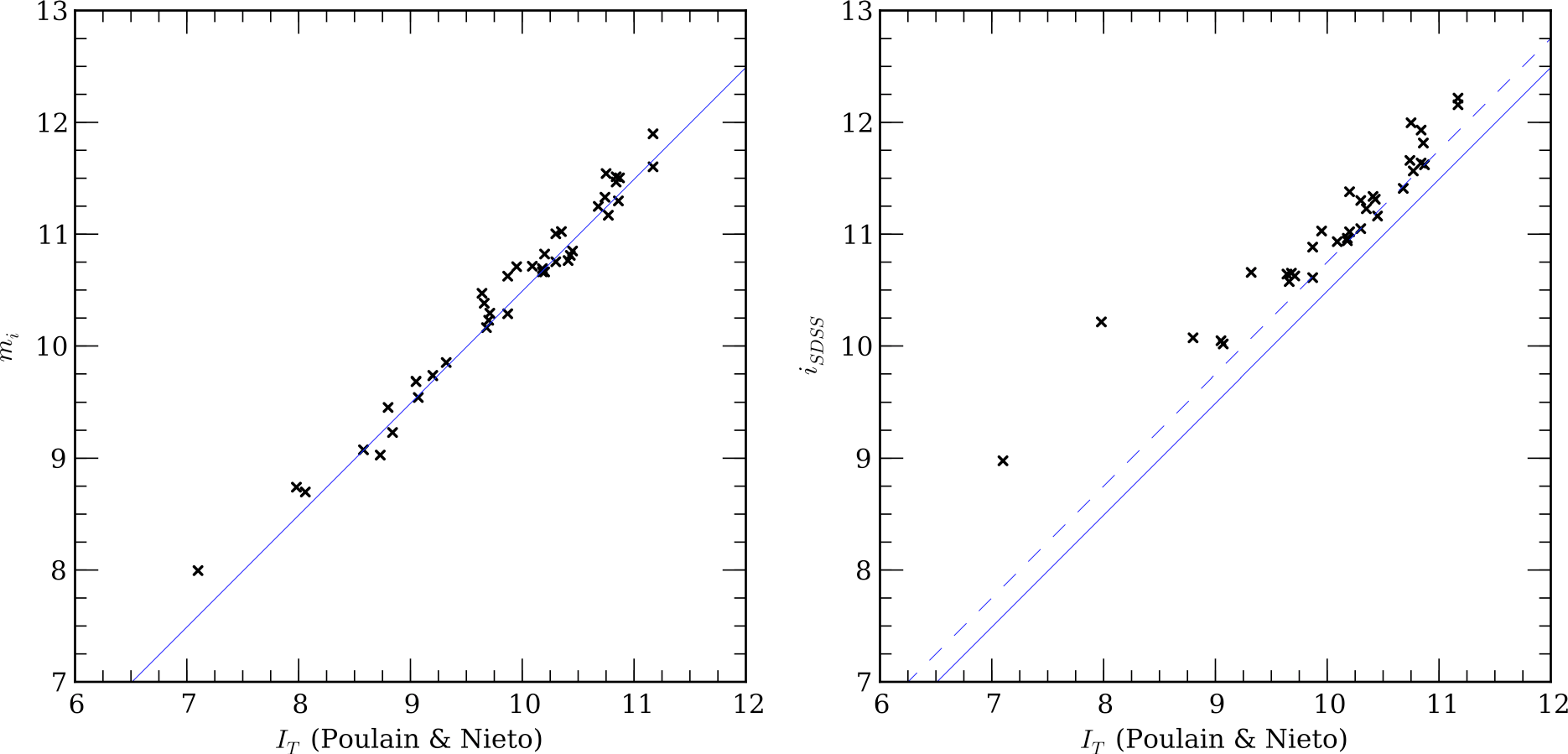}
\caption{\small The same as Figure \ref{petro_UT} for $i$ and Johnson $I$.  Symbols
and lines are the same as Figure \ref{petro_UT}, using $i$ to $I$ conversion factors.
The RC3 lacks Johnson $I$ measurements, so only data from Poulain \&
Nieto is displayed.  The dispersion is 0.14 mags for our luminosities, 0.22 mags for
SDSS Petroisan magnitudes.  There is some indication that SDSS Petroisan $i$ values
underestimate the total luminosity by 10 to 20\%.
}
\label{petro_IT}
\end{figure}

For comparison, we display the total apparent luminosities ($m_g$, uncorrected for
Galactic extinction) versus SDSS DR12 Petrosian, RC3 $UBV$ and Poulian \& Nieto
$UBVRI$ photometry in Figures \ref{petro_UT}, \ref{petro_VT}, \ref{petro_RT} and
\ref{petro_IT}.  The luminosity measure of choice for the SDSS project is the
Petrosian magnitude defined as an aperture magnitude where the aperture size is given
by the Petroisan index (Petrosian 1976), a metric measure based on the shape of the
galaxy's luminosity profile.  The Petrosian magnitude, as it relates to the total
galaxy luminosity is well-studied (see Graham \etal 2005 and references therein).
Its primary deficiency is that it uses circular apertures, that will introduce extra
sky noise for elongated galaxies, and the fact that different profile shapes will
lead to variation between the true total luminosity and the Petrosian flux.  For
$r^{1/4}$ shaped profiles (that is a 1st order shape for a majority of ellipticals),
the difference is 0.20 mags (approximately 27\%).  And this value approaches 100\% as
the galaxy becomes smaller in angular size (Blanton \etal 2001).

The left panel of Figure \ref{petro_UT} displays the comparison between this studies
$u$ total luminosities ($m_u$) and the RC3 (103 galaxies in common) plus the PN94
UBVRI sample (Poulian \& Nieto 1994, 40 galaxies in common, red symbols).  The blue
line represents the conversion from Johnson $U$ to SDSS $u$ assuming a mean $U-B$
color of 0.48 and mean $B-V$ color of 0.97 (resulting in a $u-U=0.87$, see Jordi
\etal 2006).  The dispersion around the blue line is 0.40 mags, that is consistent
with the calculated errors in our $u$ fluxes and the quoted RC3 errors.

The right panel of Figure \ref{petro_UT}, a plot of RC3 and PN94 $U$ magnitudes
versus DR12 Petrosian $u$ magnitudes ($u_{SDSS}$), displays a different picture.  All
the quoted SDSS errors are much smaller than our errors, but relationship between RC3
$U_T$ and $u_{SDSS}$ has a much higher scatter.  The solid blue line in Figure
\ref{petro_UT} again reflects the expected $u-U$ conversion.  The dashed blue line
represents the expected relationship if the SDSS Petrosian magnitudes are 80\% the
total luminosity.  The relationship is similar to our total luminosities; however,
the scatter in the SDSS Petroisan magnitudes is 0.80 mags.

Similar comparisons are found in Figures \ref{petro_VT}, \ref{petro_RT} and
\ref{petro_IT}.  As the RC3 only contains $UBV$ magnitudes, the PN94 sample are used
for comparison in $R$ and $I$.  As in Figure \ref{petro_UT}, the blue line represents
the standard conversions from SDSS filters to the Johnson system for the mean color
of an elliptical (Jordi \etal 2006).  The dashed lines represent those conversions if
the Petrosian magnitudes contain 80\% of the flux.  The dispersions for our
magnitudes are 0.15 for $g$, 0.14 in $r$ and 0.14 in $i$.  The only notable
difference is that the SDSS $i$ magnitudes are significantly deviant to Poulain \&
Nieto values for unknown reasons.

\begin{figure}[!ht] 
\centering 
\includegraphics[scale=0.75,angle=0]{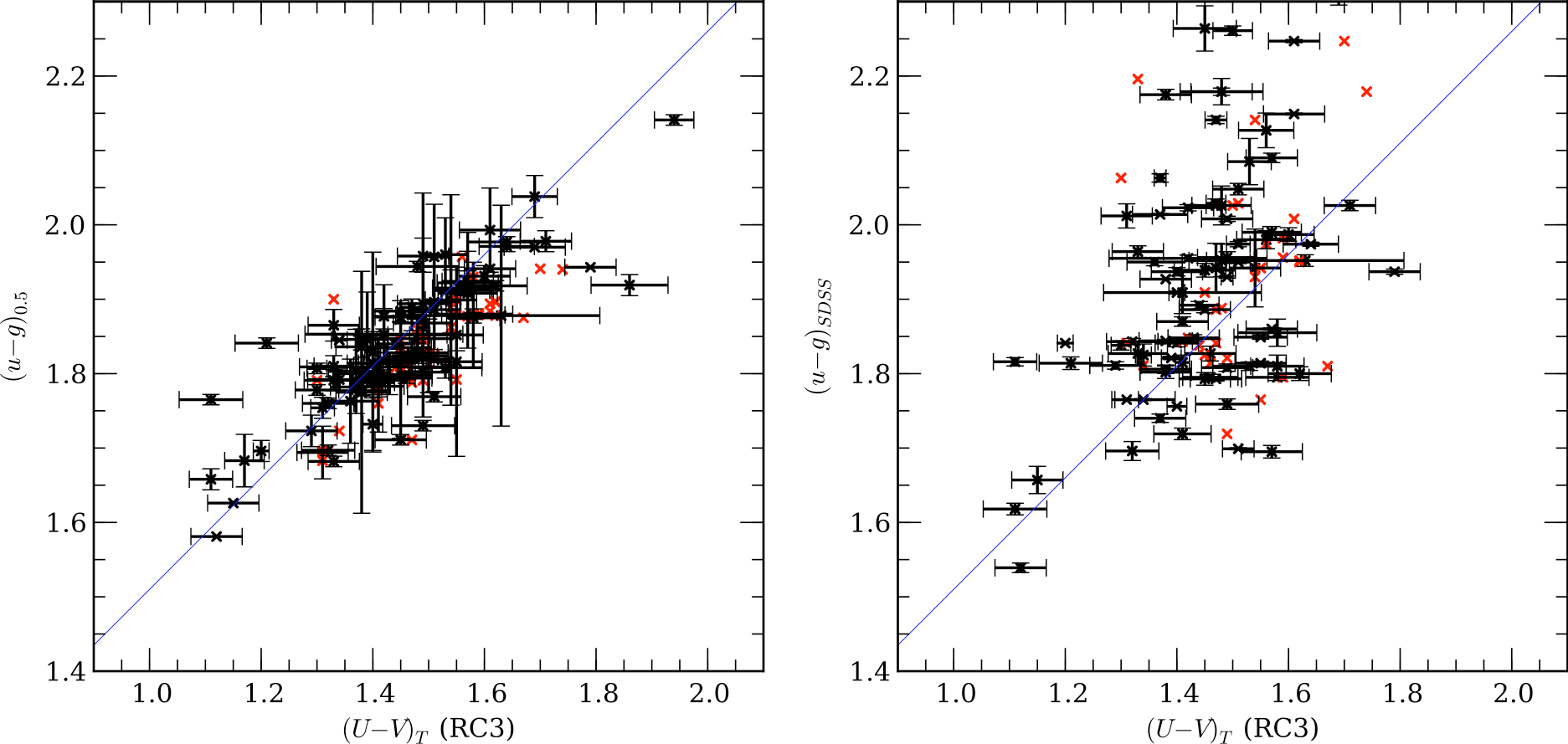}
\caption{\small A comparison between Johnson $U-V$ colors for ellipticals and our
half-light $u-g$ colors ($(u-g)_{0.5}$) and DR12 Petrosian colors.  The symbol colors
are the same as Figure \ref{petro_UT}.  The blue line displays the expected
conversion from $U-V$ to $u-g$.  The dispersion around the conversion line is 0.09
with a small offset of 0.10 mags.  The DR12 Petrosian colors are, unsurprisingly, a
poor match to $U-V$ colors.  This is primary due to the fact that different filters
will have different Petrosian aperture sizes resulting in a mismatch in fluxes.
Reliable SDSS colors use aperture luminosities store in the DR12 database luminosity
profiles.
}
\label{UV_ug}
\end{figure}

Lastly, we present a comparison of Johnson $U-V$ colors to SDSS $u-g$ in Figure
\ref{UV_ug}.  The agreement between our half-light $u-g$ colors and RC3 $U-V$ is
good.  The blue line displays the expected linear conversion between $u-g$ to $U-V$
(a difference of 0.75); however, the conversion appears to have a slight color term.
Otherwise the agreement between the RC3 and this study is excellent.  The right panel
displays the $u-g$ color from the DR12 database and clearly displays no correlation.
This is due to the method of defining SDSS magnitudes, correct colors require
extracting the raw image profiles from the DR12 database and perform aperture colors
on the intensity profiles.  In other words, this is not a statement on the quality of
the SDSS data, but rather a cautionary tale for users of the SDSS datasets.

\subsection{2MASS}

The Two Micron All Sky Survey (2MASS) was a ground-based, all-sky survey in three
near-infrared filters; $J$ (1.2$\mu$m), $H$ (1.7$\mu$m), and $K$ (2.2$\mu$m).  The
detectors were 256x256 HgCdTe arrays with 2 arcsecs pixels, but using a freeze frame
scanning technique produced final images that were 1 arcsec in plate scale with a
total of 7.8 secs of exposure time per pixel.  Galaxy photometry from 2MASS is
presented in the 2MASS Extended Source Catalog (Jarrett \etal 2000) and the 2MASS
Large Galaxy Atlas (Jarrett \etal 2003).  As noted in Schombert (2011), the
photometry in those catalogs suffered from systematic bias owing to an erroneous sky
determination routine.  While those quoted photometry measurements are typically 0.33
mags too faint, their values are still quoted in the NED database.

\begin{figure}[!ht]
\centering
\includegraphics[scale=0.75,angle=0]{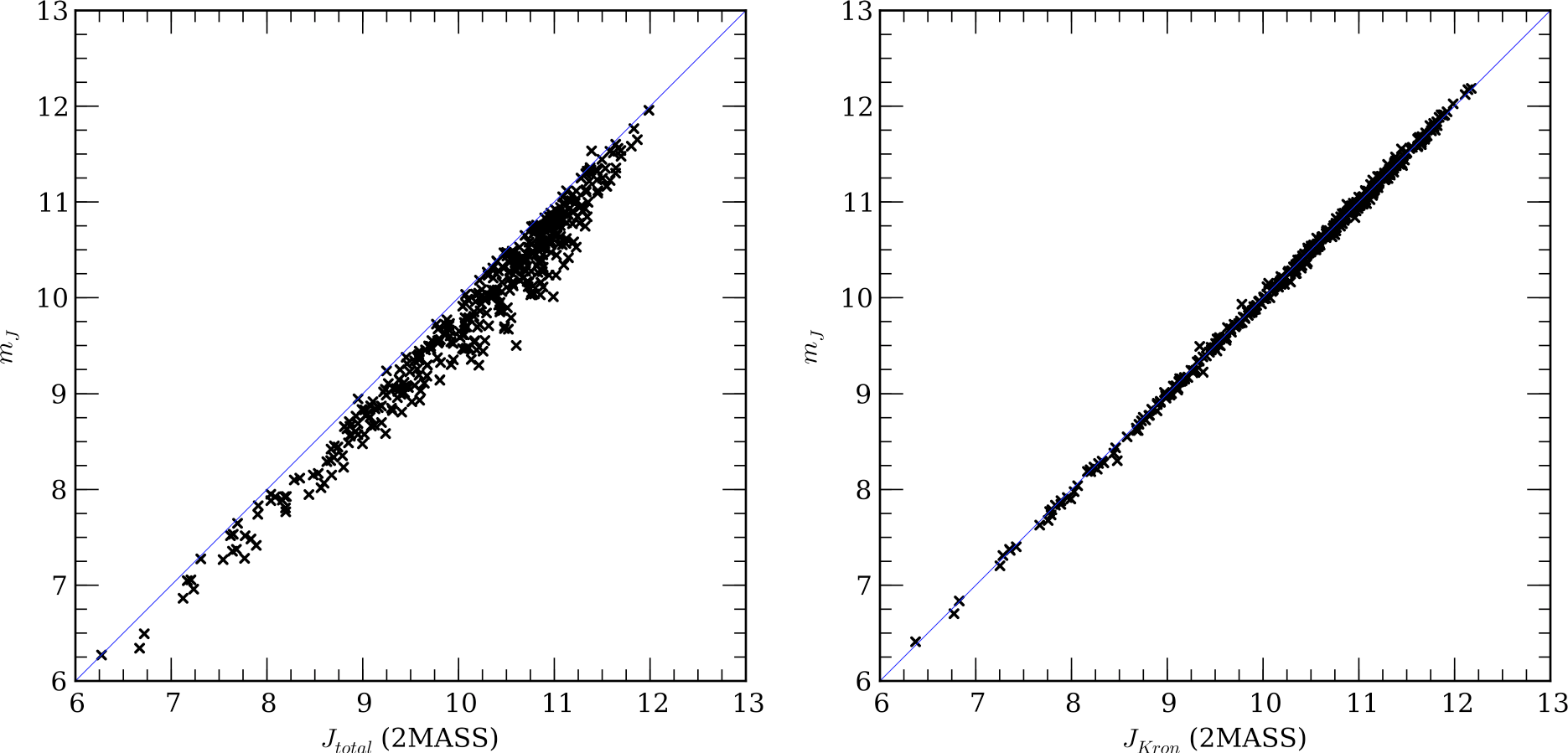}
\caption{\small A comparison between our total $J$ magnitudes ($m_J$) and the total
magnitudes from the 2MASS Large Galaxy Atlas (Jarrett \etal 2003).  The unity line is
shown in blue.  The large scatter and offset reflect a known problem for 2MASS sky
subtraction in the LGA (see Schombert 2011).  The 2MASS $J$ Kron magnitudes (right
panel) use aperture values that are in good agreement with our values, although the
expected 20\% offset is missing.  }
\label{2mass_total}
\end{figure}

A more accurate value for total luminosities in $JHK$ are the Kron magnitudes from
Jarrett \etal (2003).  Figure \ref{2mass_total} displays the comparison with our
luminosities at $J$ and the 2MASS Extended Source photometry (left panel) and Kron
$J$ magnitudes (right panel).  The $J$ total magnitudes still display the irregular
offset, but the Kron magnitudes are in good agreement with this study.  This would
indicate that the Kron magnitudes from Jarrett \etal (2003) are much closer to a
total luminosity than normally expected from an aperture value and, importantly, do
not require the 80\% correction typical for Kron magnitudes.

\subsection{\it Spitzer}

The {\it Spitzer} Space Telescope was the last of NASA's Great Observatories program
launched in 2003 and still in operation.  The 0.85m Ritchey-Chretien is sensitive
from 3.6 to 160$\mu$m.  It's primary imaging instrument is the InfraRed Array Camera
(IRAC) which is a four channel device (3.6, 4.5, 5.8 and 8.0$\mu$m) using 256x256
arrays that have a plate scale of 1.2 arcsecs per pixel.  Long exposures for
extragalactic objects involved a number of differing dithering schemes that reduce
the plate scale by 1/2 for the final image products, although the typical FWHM PSF
was 1.7 arcsecs.

The details and difficulties of analyzing {\it Spitzer} images were described in
Schombert \& McGaugh (2014).  Perhaps the most salient difference between optical and
Spitzer imaging is the sharp increase in the number of point sources, not associated
with the galaxy of interest, visible in each frame (see Figure 2 of Schombert \&
McGaugh 2014). The number of point sources in the 3.6$\mu$ image is a factor of 10
greater than the number in the $V$ image, although this is expected from early
Spitzer number counts (Fazio \etal 2004).  Aggressive cleaning is required in all the
elliptical images, and replacement of the masked pixels was a critical step in the
surface and aperture photometry.

\begin{figure}[!ht]
\centering
\includegraphics[scale=0.75,angle=0]{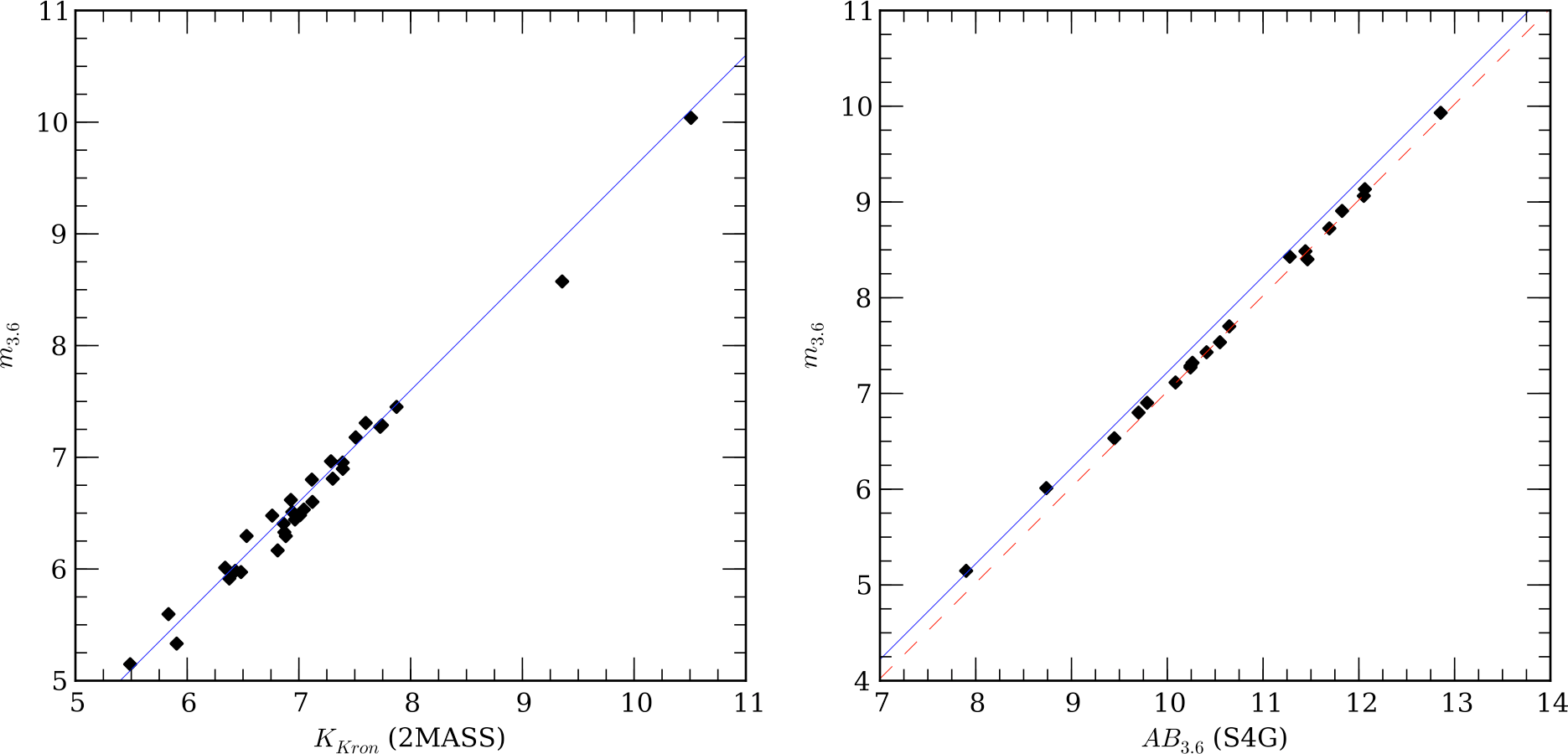}
\caption{\small A comparison of our total 3.6$\mu$m magnitudes (on the Vega system) and
2MASS $K$ and total 3.6$\mu$m magnitudes from the S$^4$G project (Sheth \etal 2010).  The blue
line in the left panel displays a difference of $K-3.6 = 0.4$ (Schombert \& McGaugh
2014).  The blue line in the right panel display the conversion from Vega to AB system of
2.779.  The dotted red line represents the Kron magnitude of 80\% the total flux.
The agreement with 2MASS $K$ magnitudes is within the photometric errors and
confirms the expected mean $K-3.6$ color of ellipticals to be 0.4.  The agreement
with S$^4$G magnitudes is also excellent on the assumption that S$^4$G luminosities
are measured through Kron apertures.
}
\label{spitzer_compare}
\end{figure}

Figure \ref{spitzer_compare} displays a comparison of our $m_{3.6}$ luminosities and
2MASS $K$ Kron magnitudes and total 3.6$\mu$m luminosities from the S$^4$G project
(Sheth \etal 2010).  For the limited number of galaxies in common, the agreement is
good, better for the S$^4$G magnitudes than the 2MASS $K$ values.  The unity line in
the left panel assumes a $K-3.6$ color of 0.4 for ellipticals and the scatter is
consistent with the short exposure values of 2MASS.  The greater depth to the {\it
Spitzer} images is reflected in the lower scatter between our luminosities and the
S$^4$G luminosities using the same image files.

\section{Colors}

\subsection{Aperture Effects}

While the relative flux through two filters (i.e., a color) is a fundamental
observable for galaxies, the manner in which we assign that value varies with
photometric technique.  Our earliest photometric studies on ellipticals used circular
apertures fixed to the cathode photometer technology of the time.  This had the
disadvantage of forcing a circular shape on a typically non-circular object plus
foreground and background objects are included in the aperture values.  Multiple
aperture values allows the construction of a curve-of-growth to assign a total
luminosity in each filter, which is then converted into a total color.  The RC3 (de
Vaucouleurs \etal 1991) has long been the leading repository of photoelectric
measurements for the brightest galaxies in the sky, and the ${UBV}_T$ system is the
standard for comparison of galaxy colors.

The development of areal detectors (e.g., CCD cameras) allowed for a more
sophisticated 2D surface photometry analysis of galaxies.  Foreground stars and
background galaxies could be masked from the galaxy flux.  Elliptical apertures can
be applied to the galaxy shape, minimizing the contribution from sky noise and
eliminating aperture corrections.  Isophotal and metric values could be deduced from
the galaxy's surface brightness profile and color gradients could be analyzed in
greater detail.

A wealth of photometric information makes the comparison of luminosity and structural
quantities problematic between various studies.  Metric magnitudes can be adopted,
but the elliptical shape of the galaxy has to be defined as well as the metric
diameter used.  With respect to color, Scodeggio (2001) demonstrates that effective
sizes vary with wavelength such that isophotal colors produce highly erroneous
results.  The most common color analysis for ellipticals uses total luminosities
based on the extrapolated curves-of-growth; however, while total luminosities are
best obtained by this technique, large errors in the outer apertures introduce
unneeded errors in the total color.

The different types of determinations of galaxy colors are presented in Bernardi
\etal (2005) for an analysis of SDSS luminosities of early-type galaxies.
Historically, galaxy color determination has fallen into five categories: 1)
effective radius color, 2) fixed angular size color, 3) Petrosian colors (Stoughton
\etal 2002), 4) model colors and 5) half-light colors.  The first two types of color
determination use fixed apertures that produce the lowest error differences as the
flux is measured in two filters through the same spatial pixels per galaxy.  However,
a fixed angular size only proves reliable if the target objects are at similar
distances so that the aperture measures some constant metric size.  An aperture
defined by the effective radius (e.g., based on a r$^{1/4}$ or S\'{e}rsic fit to the
surface brightness profile) has the advantage of being distance independent and
reflects a metric size that scales with the total size of the galaxy.  However, often
fitting functions provide different values with different filters, so the measured
color may vary widely from filter to filter pair making comparison between colors
problematic.

The Petrosian magnitudes are luminosity determinate of choice for the SDSS project.
They are highly reproducible and are excellent measures of the total luminosity of a
galaxy (capturing 80 to 90\% of the total flux).  However, this system is primarily
designed for small angular sized galaxies and its reliability for large galaxies
decreases (see \S2.3).  While the total luminosities of large galaxies are consistent
from faint to bright systems, the errors increase with larger galaxy size and make
the colors deduced from Petroisan colors less than optimal.

The use of model fit colors is popular amongst SDSS studies of galaxy colors
(Bernardi \etal 2003).  Here a r$^{1/4}$ or exponential profile is fit to the
isophotes (Strauss \etal 2002) and these structural values are used to deduce a total
luminosity.  This technique, again, produces reliable and repeatable luminosities for
small galaxies (i.e., typically high redshift) and agrees well with Petrosian
magnitudes.  But the errors introduced with the fitting functions compounds with
comparison between filters (again, see \S3.1).

As concluded by Bernardi \etal (2005), fixed aperture colors have the highest
reliably and repeatability.  As all our galaxies have large angular extent, high S/N
images, and we have adapted a similar procedure herein.  Using the technique as
outlined in \S3.2, an angular size is defined by some fractional luminosity
(typically the half-light point) in a particular filter.  Then, that same aperture is
examined in all of the filters to produce the colors.  This procedure results in the
lowest RMS values for color relations, such as the color-magnitude relation.  For our
study, we have used the half-light radius as defined by the $J$ band images.  While
other fractional colors could have been selected (see below), experimentation
demonstrated that, given the variable S/N between the datasets, that larger radii
results in higher errors in the {\it GALEX} images and smaller radii became less
representative of the total color of an elliptical owing to effects from color
gradients.

\begin{figure}[!ht]
\centering
\includegraphics[scale=0.75,angle=0]{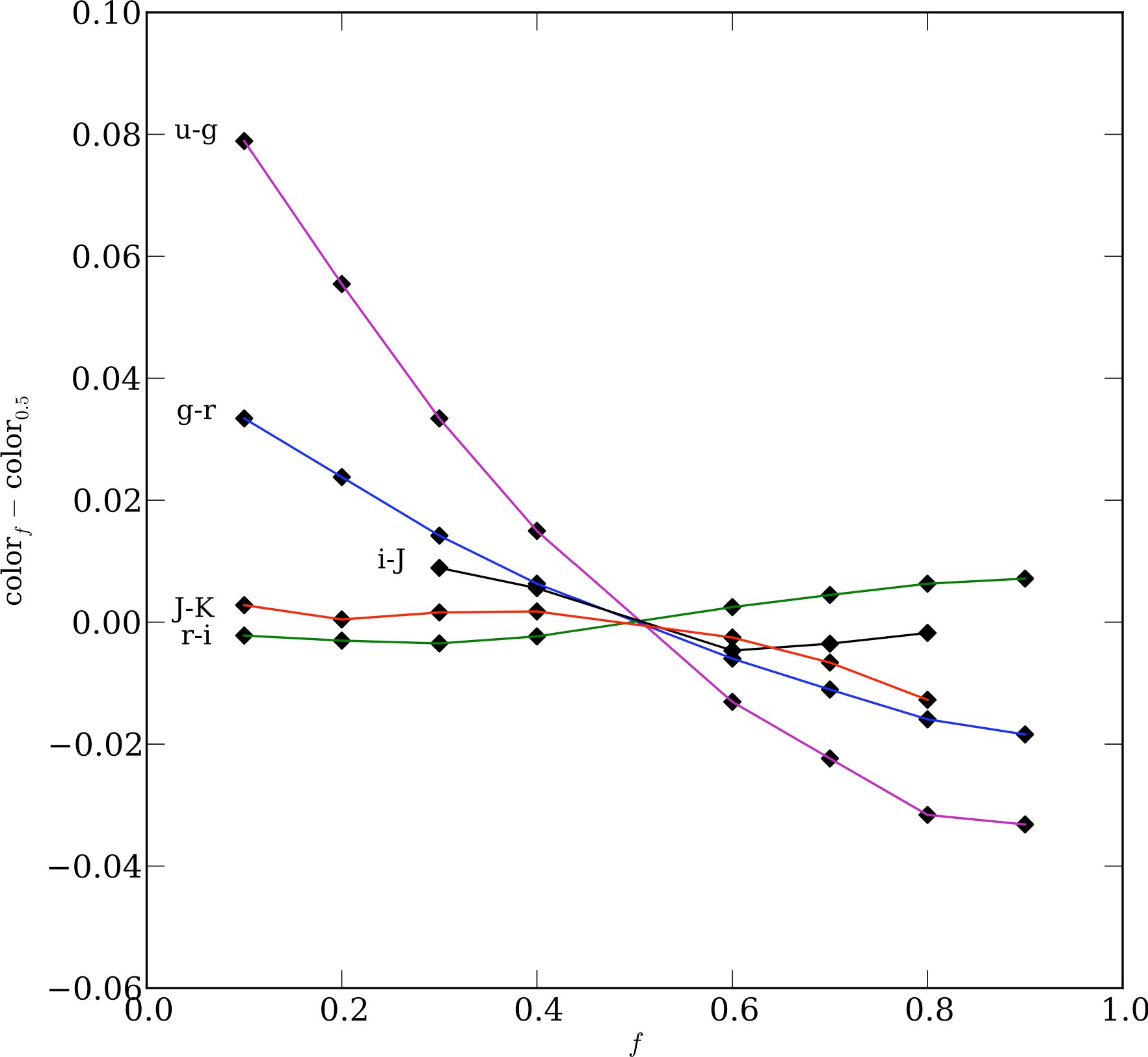}
\caption{\small The difference in mean galaxy colors as a function of fractional aperture.
Using the total $J$ flux values, a fractional light radius is defined for each
galaxy.  Colors are measured through those metric values for all filters from $u$ to
$K$ ({\it GALEX} and {\it Spitzer} data not shown).  The mean difference between the color
at each fractional radius and half-light color (color$_{0.5}$) is plotted as a function
of the fractional radius.  This is a aperture flux measure of the known effect of
color gradients in ellipticals (Michard 2000).  The color gradient effect is strongest for
the bluest filters, although typically less than 0.04 for most blue filters and
negligible for near-IR colors.
}
\label{frac_colors}
\end{figure}

It also must be remembered that an aperture color is a 2D slice through a 3D object.
Thus, any aperture will produce a color that is a 3D cylinder from the galaxy's
envelope through its core.  Since most ellipticals display color gradients, thus even
a fixed aperture color will be the composite of the stellar populations in the core
and envelope, which display different spectral signatures.  This will be mitigated by
the fact that stellar density increases towards the core, such that smaller apertures
are increasingly dominated by core luminosities.  But, any color or spectral value
based on an aperture measurement is a luminosity weighted 2D projected value that 
does not vary linearly with radius.  And interpretation of colors resulting from the
underlying stellar population requires the careful choice of an aperture size that
maximizes the light from the region of interest (i.e., core versus envelope).  

For example, a galaxy with a r$^{1/4}$ shaped profile following a Prugniel-Simien 3D
density profile (Prugniel \& Simiens 1997, Terzic \& Graham 2005) has a projected
half-light radius ($r_h$) that is slightly larger than the projected effective radius
($r_e$, about 10\% larger) and contains, obviously, 1/2 the luminosity of the galaxy.
However, of that light, 90\% is emitted by the core regions of the galaxy at less
than 1.5$r_e$.  For an aperture of 2$r_e$, 70\% of the total luminosity is measured
and 90\% of that flux comes from the inner 2.2$r_e$.  For a large aperture (e.g., a
Kron magnitude that contains 90\% of the total light of a galaxy), approximately 50\%
of the light comes from the core regions (1.5$r_e$), 40\% from the envelope region
(between 1.5 and 5$r_e$) and only 10\% from the halo (greater than 5$r_e$).  For an
typical color gradient, this will result in a blueward shift in the colors with
larger and larger apertures.

A effort to quantify the effect on colors by selecting different apertures is shown
in Figure \ref{frac_colors}.  Here the aperture colors are calculated for all the
galaxies in our sample at fractional light intervals of 0.1.  The half-light color
($f=0.5$) is used for normalization and the difference between the fractional color
and the half-light color (color$_f$ $-$ color$_{0.5}$) is plotted as a function of
the fraction of luminosity ($f$).  Due to typically blue color gradients for
ellipticals (Michard 2000), it is unsurprising to find that small apertures measure
redder colors than larger apertures.  The color difference is also dependent on
wavelength, with bluer filters display a greater fractional difference in agreement
with the finding from color gradients studies (Kin \& Im 2013).  For filters redward
of SDSS $r$, the choice of fractional light radius becomes negligible, as do their
color gradients.

The estimate of accuracy for aperture colors depends on the size of the aperture
used.  For small apertures, the errors are dominated by Poisson noise and calibration
uncertainties.  Repeatability becomes an issue for small apertures, when comparing to
other published values, due to centering errors and the unknown shape and size of the
apertures (as elliptical apertures are used in our study to minimize sky
contribution).  Thus, while internal errors may be small, comparison between studies
may develop large zeropoint differences.  

As the size of the aperture grows, the dominant component to the noise becomes the
uncertainties in the correct value for sky.  Even for the half-light aperture, the
sky noise for most of our sample contributes from 50\% of the error in the UV to 30\%
in the optical and 40\% in the IR.  Accurate flattening and pixel-to-pixel errors
also play a significant role in the error on the sky value.  For our final sample, we
estimated the error by using a robust Monte Carlo simulation for each galaxy through
all the available filters.  The simulation calculated the sky noise based on
variations between difference positions within the frame as well as the dispersion on
the mean value within a localized box.  Then, using the RMS error around each
isophotal ellipse, the simulation derives a mean error for each radius, which
reflects into the integrated aperture flux.  The error in each filter is added in
quadrature when combined with other filters to form the error on the color.  The
error bars quoted in Figures \ref{arch_galex} to \ref{spitzer_compare} are determined
by this method.  And, it is clear from inspection of these Figures, that the
luminosities and colors are reproducible from study to study, but that the formal
errors quoted by most studies are significantly out of balance with the scatter in
the unity lines, particularly those fluxes from the SDSS project.

The half-light colors (color$_{0.5}$) were determined for all the galaxies in the
sample using the same angular elliptical aperture from filter to filter.  All
the colors presented in the rest of the paper are half-light values, in all Figures and
Tables.   Total luminosities (and, thus, determination of the half-light radius) were
anchored to the 2MASS $J$ filter, as the flux from galaxies peaks around the $J$
wavelength.  The average half-light colors across all the filters in our sample are
shown in Table 2.  The Table is divided into five luminosity classes given as
multiples of the canonical cluster galaxy luminosity, $L_*$ (defined as roughly
$2\times10^{10}$ $L_{\sun}$ or $M_B = -20.6$).  Each mean color is calculated using a
jackknife technique with a 3$\sigma$ cut for outliers.  The dispersion (not the
error) on each color is shown as well as the number of galaxies in each color bin.
The global effect of the color-magnitude relation is seen in each luminosity bin with
redder colors for the brighter luminosities (see \S5).  It is important to note, for
conversion from magnitudes to fluxes, that {\it GALEX} $NUV$, SDSS $ugri$ and {\it
Spitzer} 3.6 magnitudes are in the AB system versus the Johnson magnitudes used by
2MASS $JHK$.  

Comparison to previous color studies can be problematic as techniques and aperture
sizes will vary.  Due to color gradients, varying aperture sizes can significant
alter the mean colors one would calculate and produce discrepant values from study to
study without actually being due to incorrect photometry.  For example, comparison to
past SDSS studies often displays a difference as many use the provisional SDSS system
(called $g^*$,$r^*$,$i^*$, see Bernardi \etal 2003).  This results in a color offset,
for example, $g-r$ in this study is approximately 0.035 mags redder than $g^*-r^*$
and $r-i$ is 0.070 redder than $r^*-i^*$.  In contrast, the work of Hansson, Lisker
\& Grebel (2012) finds similar $ugri$ colors as those in Table 2.   For their
brightest galaxies, they find a mean $u-g$ of 1.8, $g-r$ of 0.8 and $r-i$ of 0.4.
Zhu, Blanton \& Moustaka (2010) present a $g-i$ CMR for the red sequence that
progresses from $g-i = 1.0$ for faint ellipticals to $g-i = 1.2$ at the bright end,
which is slightly bluer than our colors over the same luminosity range.  Chang \etal
(2006) presents optical and near-IR colors for ellipticals and find that their
average $gri$ colors match our values, but their $J-K$ values are about 0.1 redder
($J-K$) than previous studies.  In addition, as we will see in \S3.4, our colors are
consistent with ellipticals being an extension of globular cluster colors.

Comparison to historical photometric samples, such as the Poulain \& Nieto samples
and the RC3, are very favorable.  The mean $U-B$ and $B-V$ colors for ellipticals are
0.51 and 0.95 from the photoelectric catalogs.  This corresponds to a mean $u-g$ of
1.82, identical to our $L_*$ value.  From Poulain \& Nieto, we have a mean $V-R$ of
0.59 and $R-I$ of 0.67 that corresponds to a $g-r$ of 0.84 and a $r-i$ of 0.43.
Again, agreement with our bright elliptical colors.  In addition, our $J-K$ and $g-K$
colors agree with large galaxy samples from 2MASS and SAURON (Falc{\'o}n-Barroso
\etal 2011).  While this comparison is crude, it does illuminate the growing
consistency between photometric studies from large sky surveys.

Interpretation of color will require comparison to the various SED models in the
literature.  While most of these SED codes have converged to produce similar results
for identical inputs (mostly owing to improved stellar tracks, see Schiavon 2007), at
the far ends of our colors ({\it GALEX} and {\it Spitzer}) exotic stellar populations
play an increasing difficult role.  For example, intermediate aged populations of
asymptotic giant branch (AGB) stars have a significant contribution to the IR colors
(Schombert \& McGaugh 2014).  Their luminosities are bright enough such that small
variations in their individual numbers make measurable changes to even integrated
colors (Schombert \& McGaugh 2014).  In the UV, the contribution from hot horizontal
branch (HB) stars plays an important role, especially in envelope colors where low
metallicities are expected (Yi 2011).  This will be discussed further in \S6.

\begin{figure}[!ht]
\centering
\includegraphics[scale=0.75,angle=0]{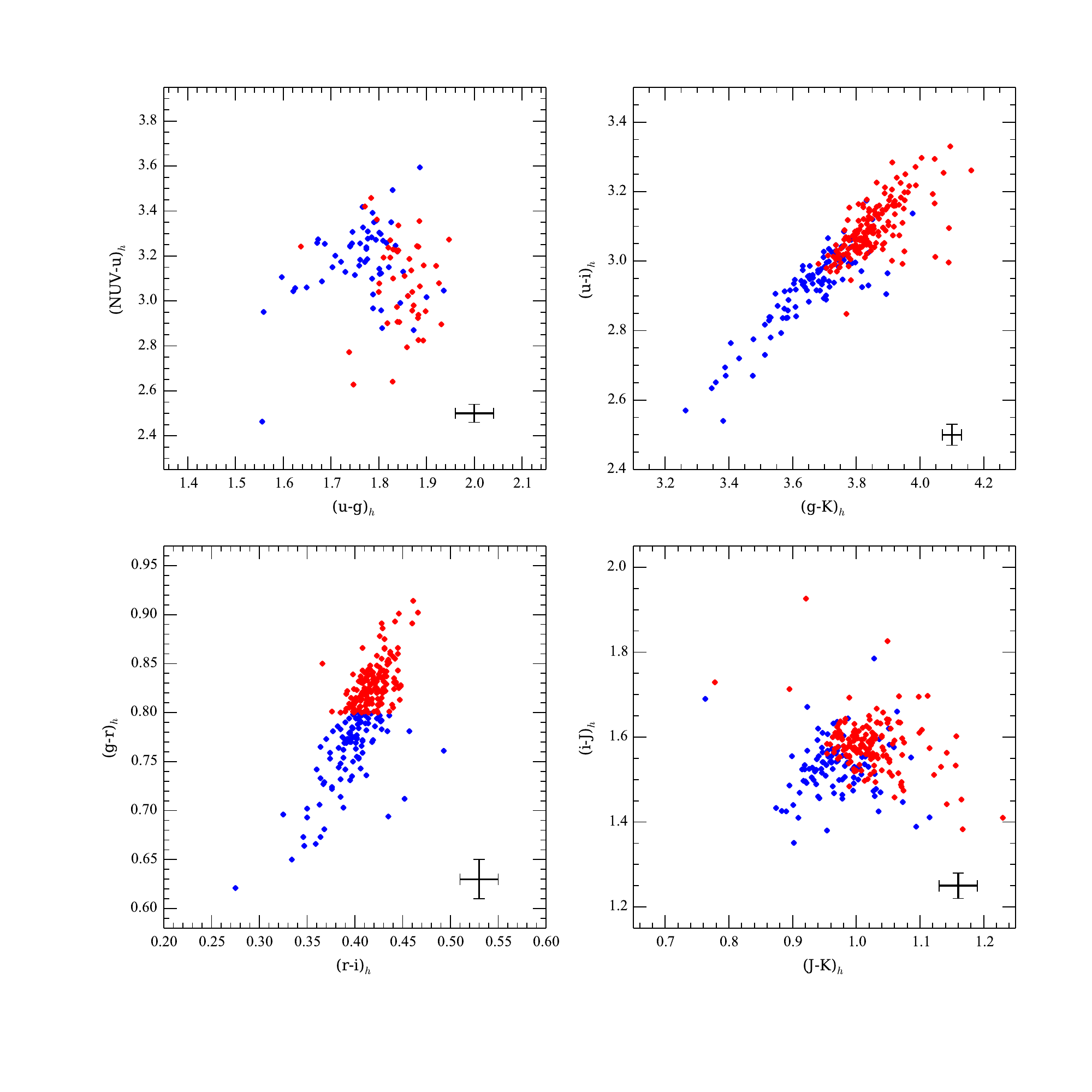}
\caption{\small Four two color diagrams are displayed, three with connecting colors
(e.g., $NUV-u$ versus $u-g$) and one with a large baseline ($u-i$ versus $g-K$).  All
colors are based on the half-light radius defined in $J$. The coherence in color is
very clear in the sample (i.e., the bluest ellipticals are blue in all colors and
vice-versa).  To see this effect more clearly, the sample is divided by $g-r$=0.80
and designated by symbol color).  Photometric errors dominate the scatter in the
color-color relations, indicated that the stellar populations in ellipticals vary in
a smooth and uniform fashion within the sample, although age and metallicity changes
are mostly parallel to the color-color sequences.  The $NUV-u$ colors differ in
behavior from the optical and near-IR relationships, and will be discussed in \S6.
}
\label{6color_split}
\end{figure}

\subsection{Two-Color Relations}

Four two-color diagrams (of the possible 45 color combinations) are shown in Figure
\ref{6color_split}.  Typical error bars are shown in the bottom right corners.  The
trend for positively correlated colors (colors are correlated from blue to red) is
seen in all colors except for $NUV-u$ (see below).  The sample is divided into blue
($g-r < 0.8$) and red ($g-r > 0.8$) galaxies, and each panel demonstrates that
elliptical colors are, in general, coherent from filter to filter (i.e., red galaxies
are red in all filters).  In addition, outliers are usually outliers across all their
colors with the probable explanation that some contamination is in the galaxy itself
(e.g., an embedded foreground star).  Other outliers may be due to undetected strong
emission lines or simply a flaw in the calibration or reduction pipeline.  Their
rarity did not warrant extensive investigation as they are not relevant to the
averaged results.

Correlations between colors have the lowest scatter for widely spaced filters, mostly
because filters close in wavelength (e.g., $J-K$) have less dynamic range (the slope
of the galaxy SED varies little over small changes in wavelength except near the
4000\AA\ break) and photometric errors play an increasing role.  The color of an
elliptical is correlated with its luminosity (the color-magnitude relation, see \S5)
and distribution of colors reflects a true range in internal color due to differences
in the stellar populations (i.e., mean metallicity and age), not because of
correlated errors.  There is no evidence of recent star formation, signaled by
$NUV-r$ colors less than 5.5, (Schawinski \etal 2007), although very small amounts of
recent star formation (less than 1\% the total mass, Rakos \etal 2008) would be undetectable in
integrated colors (2D color mapping would be more useful for this issue).

To repeat the discussion in \S3.1, the half-light colors are defined by the metric
radius from half the total luminosity as determined in the $J$ band.  This metric
elliptical aperture is then applied to all the other images (regardless of whether is
the same half-light point).  The resulting colors are thus measured through the same
effective apertures, although the angular size will vary from galaxy to galaxy in
proportion to their total luminosity.  The half-light radius is dominated
(approximately 90\%) by the light from the regions inside an effective radius
($r_e$).  This has the advantage of minimizing effects owing to color gradients (whose
effects can be estimated from Figure \ref{frac_colors}) and also makes comparison to
spectral indices more realistic as those values are highly skewed by surface
brightness and represent the spectral values of the very inner core regions.  The low
scatter in the two color diagrams are also enhanced since colors from the core
regions are produced by stellar populations whose age and metallicity are more uniform
(presumably the oldest in age and highest in [Fe/H], although this can be tested).
Thus, the variation in the two-color diagrams will reflect the global change in age
and metallicity as due to total galaxy mass rather than artifacts due to aperture
effects.

\begin{figure}[!ht]
\centering
\includegraphics[scale=0.7,angle=0]{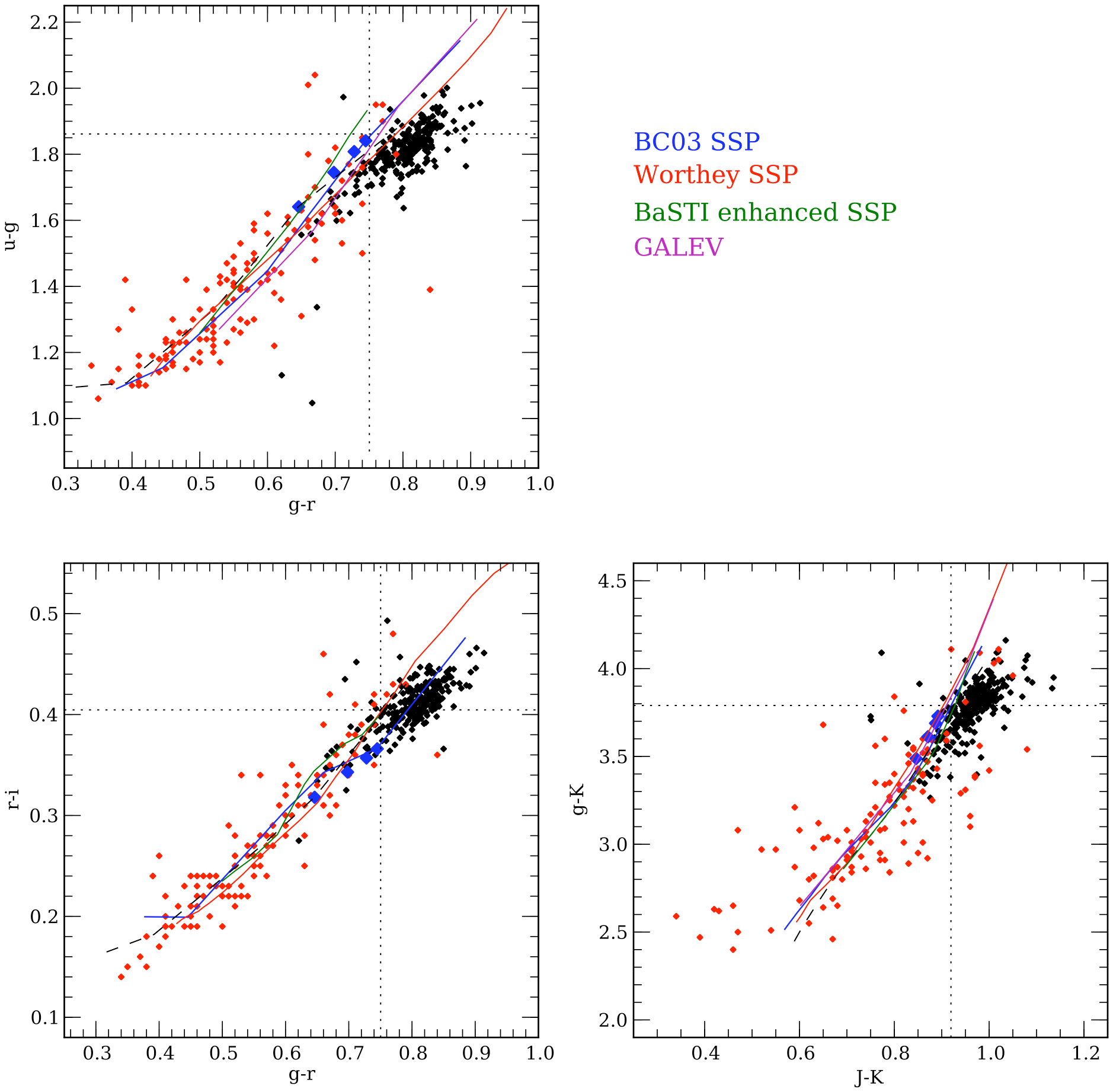}
\caption{\small Optical and near-IR two color diagrams for LMC, SMC and M31 globular
clusters (red) and ellipticals (black).  The optical photometry is from Peacock \etal
(2011), the UV and near-IR photometry is from Galleti \etal (2009).  The globulars
are restricted to those with $E(B-V) < 0.2$ and $m_g < 16.8$.  The globulars range in
[Fe/H] from $-$2.5 to solar and, if they follow the same age-metallicity scheme as
the Milky Way, LMC and SMC, then they range in age from 12 to 7 Gyrs old (Pessev
\etal 2015).  The dotted line is the solar metallicity value from globular cluster
colors.  The ellipticals and globulars overlap at the bluest ellipticals and the
reddest globulars (approximately $g-r=0.7$), strongly suggesting that the reddest
ellipticals have greater than solar metallicities, as age effects would be inadequate
to explain the reddest optical and near-IR colors.  Although the connection between
globulars and ellipticals is clear, the slope of the color-color relationships are
significantly different suggesting a mix of stellar populations in ellipticals beyond
a simple stellar population of single age and metallicity.  Also shown are four
models for 12 Gyrs SSP's and the dashed line is a BC03 5 Gyrs model (see discussion
in text).  The large blue symbols indicate the BC03 SSP values for a solar
metallicity population at ages of 5, 8, 10 and 12 Gyrs, in order of increasing color.
}
\label{globs_two_color}
\end{figure}

Historically, elliptical colors have been compared to globular clusters (Burstein
\etal 1984) due to their first-order appearance as red, non-star-forming objects.
This comparison is shown in Figure \ref{globs_two_color} where the photometry of LMC,
SMC and M31 globular clusters is compared to our elliptical sample.  The $ugri$
photometry comes from Peacock \etal (2010), a sample of 416 clusters.  The $NUV$ and
$JHK$ photometry is from Galleti \etal (2009), a subset of 225 M31 clusters.  For
accuracy, only those clusters with $E(B-V) < 0.20$ and calculated [Fe/H] values
(determined from Lick/IDS [MgFe] index values calibrated by MW globulars, see Galleti
\etal 2009) were used in our comparison.  In addition, only clusters with $V$ mags
greater than 16.8 (i.e., typical photometric errors less than 0.05) were included in
the plots and analysis.  This resulted in a comparison globular sample of 127 clusters.

As can be seen from Figure \ref{globs_two_color}, the ellipticals are a clear
extension to the globular cluster sequence from the near-UV to the near-IR.  The
large scatter in the globular cluster colors is not due to a range in ages, but
rather to the stochastic effects from variations in the number of high luminosity
stars within the cluster (Bruzual 2002).  The LMC, SMC and M31 globulars range in
[Fe/H] from $-$2.5 to solar, the solar metallicity value is indicated in each
two-color diagram.  The overlap in color suggests that the bluest ellipticals have
mean [Fe/H] values between $-$0.5 and solar, while the reddest have super-solar
metallicities (assuming that metallicity, rather than age, is the primary determinate
for global color).  If the M31 cluster system follows the MW, LMC and SMC globulars,
than the lowest metallicity globulars have ages around 12 Gyrs while those more
metal-rich than [Fe/H] = $-$0.5 are between 7 to 9 Gyrs old (Pessev \etal 2015),
which is not significantly younger than ellipticals.  Thus, the reddest ellipticals
must have greater than solar metallicities as age effects, by themselves, will not
reach those regions of the color-color diagrams.  In fact, younger ages for
ellipticals will push them blueward in the two-color diagrams, requiring even higher
metallicities to explain their red colors compared to globulars.

The relationship between colors are linear to within the photometric errors for both
the globulars and ellipticals, but with slightly different slopes.  While the bluest
ellipticals overlap in color space with the reddest globulars, the slope of the
ellipticals color-color relationships are slightly shallower than the ones found for
globulars.  This is true even though globulars span a much larger range in color at
every filter.  The shallower slopes are in the direction that the brighter
ellipticals have bluer shortward colors (i.e. bluer $u-g$ for an expected $g-r$ color
extrapolated from globulars), and the effect is also stronger in the bluer colors
(i.e. the slopes are shallower when comparing the near-blue colors) but barely
noticeable in the near-IR colors.

This difference in color-color slopes was anticipated for ellipticals based on narrow
band colors (see Schombert \& Rakos 2009) and reinforces the important result 
that ellipticals are not composed of a single age and metallicity population like a
globular cluster.  While this may be an obvious statement, based on any standard
scenario for the star formation history of a galaxy, the stellar population
characteristics of globulars and ellipticals (e.g. colors, spectral indices) are so
similar that a detectable difference has been difficult to quantify (Burstein \etal
1984).

A composite stellar population in galaxy derives from the assumption that any
extended duration in initial star formation will produce a stellar population
reflecting the duration of initial star formation in mean age and the chemical
evolution of the stars during the star formation epoch as metallicity variations.
Even if this initial burst of star formation is rapid (i.e., a very small spread in
age), there will still be significant chemical evolution (Matteucci \& Tornambe 1987)
which, in turn, produces a range of internal colors such that low metallicity
populations are typically bluer than high metallicity populations (Bruzual \& Charlot
2003).  Certainly an internal population difference is expected simply due to the
strength of color gradients detected in ellipticals (Kim \& Im 2013) that signals the
expected formation differences between halo and core stellar populations under the
canonical Eggen, Lynden-Bell \& Sandage formation scenario. 

A composite stellar population, with a range of internal metallicities, will have a
more significant color effect on high mass, high metallicity ellipticals compared to
low mass, low metallicity ellipticals.  This is due to any standard chemical
evolution scenario where some fraction of metal-poor stars exist from the earliest
epoch of star formation.  For a low mass galaxies, with less chemical evolution from
fewer SN, the spread in metallicity is small and compressed to the low end.  For
larger mass galaxies, more chemical evolution (i.e. more SN) produce higher
metallicity stars (beyond solar in many cases) where even a small fraction of older
metal-poor stars makes a significant contribution to the total color (see Schombert
\& Rakos 2009).  Thus, a low mass galaxies will have a smaller range of internal
metallicities and, therefore, a smaller range in internal color.  In contrast, a
higher mass galaxy will have a larger metal-poor component introducing bluer colors
to the redder, metal-rich component.

The change in color slope in Figure \ref{globs_two_color} could be due, solely, to a
metallicity effect in that the reddest galaxies have a low metallicity component that
makes them slightly bluer (particularly in the bluer filters due to UV contribution
from metal-poor HB stars) than expected from an extrapolation of the globular cluster
relation.  Experiments with multi-metallicity models (Schombert \& McGaugh 2014; Tang
\etal 2014) indicates the shallower slopes are what is expected from a composite
population (the value of the slope varies with assumed metallicity and age), and the
slopes should be more deviant for the bluer colors, as is seen in Figure
\ref{globs_two_color}.  We note that a standard chemical evolution scenario is all
that is required to explain the two-color diagrams, there is no need for varying ages
for ellipticals, although that possibility is not excluded from the
color-color-diagrams.

Age effects could also mimic this behavior where the reddest galaxies have a
younger mean age or a younger stellar component within an old stellar population than
the bluer galaxies (Thomas \etal 2005).  This component would need to be younger than
5 Gyrs for a noticeable change in color slope (Schombert \& Rakos 2009), and
be systematic with galaxy mass as is predicted by some hierarchical merger scenarios
(Kauffmann \& Charlot 1998; Khochfar \& Burkert 2003).  The two-color diagrams can
not distinguish between age and metallicity changes to a galaxy's colors (see \S3.3).
However, a very young age would mean that the red colors of ellipticals are due to
extremely high metallicities, which would also be difficult to understand under any
normal enrichment scenario and comparison to metallicity line indices (see \S3.3).  

In addition, a combination of age and metallicity could also explain the shallower
color slope for ellipticals in the two-color diagrams.  The degeneracy between age
and metallicity with respect to colors makes an exact analysis, using two colors,
impossible, particularly since color changes rapidly with [Fe/H] above values of
solar (see Figure 12).  But, in either the case of age or metallicity, the change in
the two color relations for ellipticals is a clear signal of a composite stellar
population.  And it is also important to note that a vast majority of the ellipticals
in this sample are redder than the solar metallicity line in all colors.  This will
be explored more fully in \S3.4, but it is important to note that the predicted
metallicities from comparison to globular cluster colors are sharply disagreement
with the deduced [Fe/H] values from spectral indices studies (see Graves \etal 2009;
Conroy \etal 2014; McDermid \etal 2015).  Spectral indices, particularly from SDSS
samples, conclude that a majority of ellipticals have less than solar metallicities
with the most massive ellipticals at values of only +0.1 to +0.2 in [Fe/H].  Figure
\ref{globs_two_color} indicates that a majority of ellipticals have greater than
solar metallicities and younger ages would only make those metallicities higher.
{\it This is the clearest example of the disparity between the deduced stellar
population characteristics of ellipticals from spectral indices and their
optical/near-IR colors (Schombert \& Rakos 2009)}.

\subsection{Comparison to SED models}

Various types of color distributions are predicted by monolithic versus hierarchical
merger scenarios (see de Rossi \etal 2007 and reference within).  Key to the
monolithic scenario is that a majority of an ellipticals stellar population is
formed in a relatively short time (less than a Gyr) and the range in color is
strictly due to a range in metallicity which, in turn, is a sole function of galaxy
mass, as steeper gravitational wells maintain more enriched gas before the onset of
galactic winds (Matteucci 2004).  This scenario often cites the well-known
color-magnitude relation as support for a metallicity interpretation to galaxy color,
where galaxy luminosity is a proxy for stellar mass.  A limited range in age to the
internal stellar populations is a requirement, one that is difficult to reconcile
with the rich history of mergers for ellipticals and the clear correlation between
$\alpha$/Fe and stellar mass (a measure of initial star formation duration).

Merger scenarios also have their drawbacks.  For example, a merger scenario would
typically predict larger scatter in the two-color relationships due to the stochastic
processes that result in an equal or unequal mass mergers with galaxies of vastly
different ages and metallicities (Chiosi \& Carraro 2002).  A merger scenario allows
for a range of ages and metallicity, as the component galaxies may have different
formation times (age) and stellar masses (metallicity).  And simulations do exist where
one can reproduce the CMR, and other color correlations, from mergers (Brooks \etal
2007).  However, they require a coordinated merger process in order to blend age and
metallicity effects into a smooth relationship of color.  In addition, mergers of
gas-rich and star forming galaxies would severely disrupt a smooth correlation of
colors with the absorption of younger stars or small bursts of recent star formation,
both of which effect integrated colors as well as disrupt smooth color gradients.

The most straight forward test of a formation and chemical evolution scenario is to
compare globular and galaxy colors to the predictions from spectroenergy distribution
(SED) models that use as few input variables as possible (Schiavon 2007).  The
primary assumptions to a SED model are 1) the initial mass function (IMF), 2) stellar
evolution tracks, 3) star formation history and 4) chemical evolution (i.e.,
population metallicity).  An idealized model of a galaxy stellar population takes a
specific amount of gas mass and produces a mass of stars following the IMF of
singular age and metallicity.  Each star (actually, a fiducial amount of the stellar
mass weighted by the IMF) is converted into a SED using a library of stellar spectra.
The population is then aged using standard stellar tracks to the current epoch (which
may vary depending on the assumed formation redshift), and the resulting stellar
SED's are summed to produce integrated colors and line indices.

A stellar population where all the gas mass is converted into stars during a single,
short duration burst of star formation, so the range in age and metallicity is
negligible, is called a simple stellar population (SSP).  These are first-order
approximations to a galaxy stellar population by assuming a limited range in internal
age and metallicity.  A much more complicated model will be needed to define an
amount of gas mass converted to stars over an extended period of time, plus a
chemical evolution model that determines the metallicity of the stars at the each
timestep.  Even for monolithic scenarios, where the star formation occurs in a burst,
there is clear observational evidence of a spread in metallicity within the
elliptical (i.e., color gradients) that will require some chemical evolution to
proceed within the short burst, even if the duration of the burst produces a very
narrow range in stellar age (a composite stellar population).  The greatest
uncertainties to SED models are the highly variable contributions from short-lived,
but high luminosity stars (such as blue HB stars in the optical and AGB stars in the
near-IR, Schiavon 2007).  Thus, comparison between competing models, each using
slightly different techniques, can be illuminating for investigating the strength of
these sub-populations on total colors and the range of predictions that these models
produce.  

For this study, we have selected four SED evolutionary models produced by BC03
(Bruzual \& Charlot 2003), Worthey (1994), BaSTI (Manzato \etal 2008) and GALEV
(Kotulla \etal 2009) for comparison to our elliptical colors.  While there are
many more SED models available in the literature, these four were selected either
because the author has detailed experience with the models from past usage to study
the narrow band colors of cluster galaxies (see Rakos \& Schombert 1995) or the
models covered the full range of wavelengths used in our dataset (plus they are well
documented and easy to access).  We make no judgement on the worth of other SED
models, nor is the primary goal of this paper to provide an extensive comparison of
all available models.

For initial comparison, all four model sets were extracted for SSPs of 12 Gyrs in age
and a range of metallicities from [Fe/H] = $-$2.5 to +0.4 (a BC03 5 Gyrs track is
also shown, see below).  These would represent the expected stellar populations in
globulars and cover the characteristics of the oldest stars in ellipticals.  These
SSP tracks, for the range of [Fe/H] values given above, are shown in Figure
\ref{globs_two_color}, color coded for each model source, along with the colors for
globular clusters and ellipticals.  The first point to note is that the variations
between the models in color space is relatively small, certainly less than the
photometric errors for either the globulars or the ellipticals from this study.
Their mapping of color into metallicity is also very similar from model to model (see
Peacock \etal 2011), which will reflect into similar colors if metallicity is the
dominant variable over mean age (e.g., globulars).  The slight differences in
two-color tracks is mostly reflected into different values for [Fe/H] assigned to the
stellar libraries and different conversions factors used to go from spectra to
colors.  To minimize the uncertainties in metallicity, each model will be tied to the
[Fe/H] values and colors of the globular clusters in the next section.  As was shown
in Peacock \etal (2011), even without these small corrections, the colors of
globulars are well reproduced by all four models.

The models fit the blue ends of the globulars two-color relationships better than the
red ends.  This is probably due to the fact that the globulars at high metallicity
(and redder color) are likely to be 2 to 3 Gyrs younger than the lowest metallicity
clusters.  Younger tracks move towards bluer colors, but to a greater degree for
bluer filters than redder.  Thus, minor age effects would predict higher deviations
from red clusters at bluer filters, which is exactly what is seen in the data.
However, the two-color diagrams are relatively useless for age determination as
changes in age from 12 to 5 Gyrs simply move the models along approximately the same
slope as the 12 Gyrs tracks, slightly bluer in the bluer filters.  For example, the
$g-r$ color of a solar metallicity 5 Gyrs SSP (the first blue symbol in Figure
\ref{globs_two_color}) is approximately 0.65 versus a $g-r$ color of 0.76 for 12 Gyrs
models of the same [Fe/H].  To entertain populations younger than 5 Gyrs, and yet
keep the colors within the mean values from Table 2, will require extremely high mean
metallicities ([Fe/H] $>$ +0.5) for ellipticals with $g-r > 0.8$, about 60\% of the
sample.  This will be more obvious with the comparison of elliptical colors to the
Lick/IDS spectral indices in \S4.

The convergence in model tracks is primarily due to the decades of feedback between
improving data (particularly in longer wavelength stellar libraries) and more
detailed modeling.  The refinement of the contributions of the late stages of stellar
evolution have dramatically decreased the discrepancies between the model tracks and
data.  However, several trends illustrate a few remaining issues between the data and
models.  For example, the poorest match between the models and colors occur for the
near-UV colors, particularly $NUV-u$ and $u-g$.  The history of modeling near-UV
colors for globulars and ellipticals is lengthy in the literature (see Yi \etal
2011).  The effects of the UV upturn to $NUV$ colors will be discussed in greater
detail in \S5 but, in brief, ellipticals are known contain a hot stellar component
that varies by galaxy mass (see Yi 2011; Rich 2005).  While metal poor populations
have a known hot component in the form of blue HB stars, even a significant
metal-poor HB population is inadequate to explain all the near-UV flux in ellipticals
(see Yi \etal 2011).  Therefore, it is common to all the models to over-estimate the
amount of flux in the near-UV filters producing model colors that are too blue from
$u$ to $r$, particularly for globular clusters.  Again, we suspect this is due to
poor conversion formula from model spectra into broadband colors.

Also salient to galaxy colors is that the SSP models are well-matched to the globular
cluster colors, but fail to reach the position of ellipticals in all the two-color
diagrams regardless of different combinations of metallicity and age.  Changes in age
simply moved the SSP tracks along the same color sequence, failing to produce the
different slopes as seen for the galaxy colors.  {\it Again, this reinforces the
conclusion from the last section that the integrated colors of ellipticals require a
composite stellar population model to explain their relationships.  And, those
population models must have metallicities greater than solar for a majority of the
ellipticals, regardless of their possible mean ages}.

\subsection{Globular Cluster Colors and [Fe/H] Calibration}

\begin{figure}[!ht]
\centering
\includegraphics[scale=0.75,angle=0]{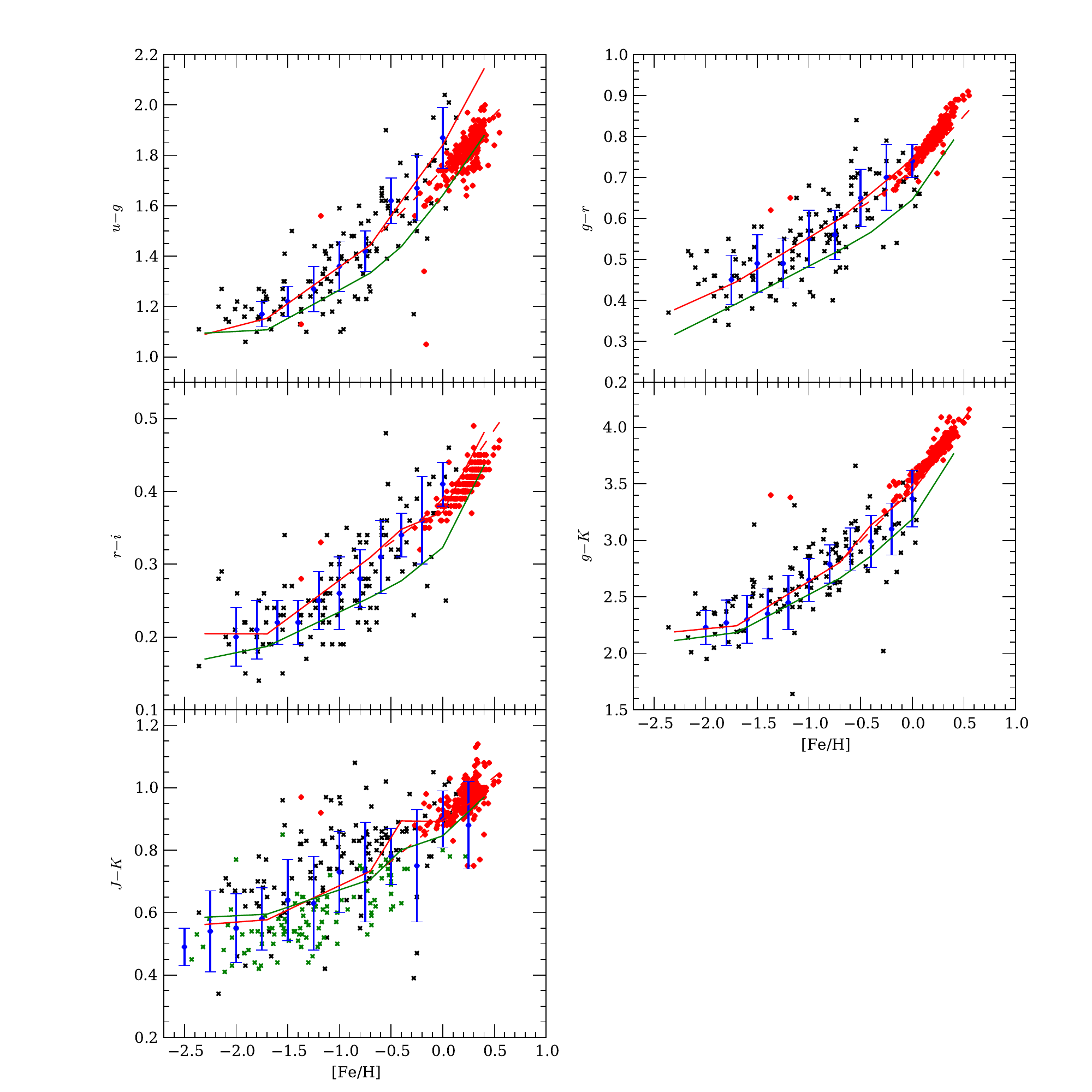}
\caption{\small The metallicity-color diagrams for five colors that the globular
and elliptical samples have in common.  The black symbols display the data for 127
globular clusters in the LMC, SMC and M31.  The blue symbols display a moving average
of the globular data with errorbars indicating the dispersion on the mean.  The red
and green tracks are the 12 and 5 Gyrs SSP models from BC03.  The dashed red line is
a 12 Gyrs multi-metallicity track.  The red symbols are the elliptical data converted
into [Fe/H] using their colors with respect to the averaged metallicity relation for
all five colors (see text).
}
\label{color_feh}
\end{figure}

The coherence of color between globulars and ellipticals suggests the possibility of
the using the correlation of color with [Fe/H] in globulars as a calibration sequence
to ellipticals, guided by SED models.  For, while age effects are difficult to
resolve using any particular color, globular metallicity versus color provides a
direct measure of the accuracy in that multiple colors can resolve average
metallicity (Peacock \etal 2011).  This is, admittedly, an oversimplified
interpretation of global colors, but does provide a first-order comparison between
approximately SSP objects (globular clusters) and ellipticals to at least test the
SSP models.  The correlation of color and [Fe/H] for globulars is shown in Figure
\ref{color_feh} for five colors.  The [Fe/H] values are determined from the [MgFe]
index as outlined by Galleti \etal (2009).  As in Figure \ref{globs_two_color}, the
globular sample is restricted to those photometric measurements from Galleti \etal
(2009) and Peacock \etal (2010) with $E(B-V) < 0.20$ and clusters with $V$ mags
greater than 16.8.  This resulting in a final globular sample of 127 clusters with
[Fe/H] between $-$2.4 and +0.1.  

Also shown in Figure \ref{color_feh} (in blue) are the moving averages for each color
using a 3$\sigma$ clipping.  The errorbars display the 3$\sigma$ dispersion from the
mean.  All the colors are well correlated with metallicity, although not linear.  The
errors in [Fe/H] ranges from 0.1 to 0.8 dex with a mean of 0.3.  If the M31 globulars
follow the same age-metallicity relation as the LMC, SMC and Milky Way, then the
clusters with [Fe/H] greater than $-$0.5 will have ages between 9 and 12 Gyrs.
Photometric error were typically 0.02 (slightly higher for bluer filters).  Given the
slopes of the relationship between color and [Fe/H], the mean errors explain all the
scatter in the relationships, primarily from uncertain [Fe/H] values.

It is possible to assign [Fe/H] values to ellipticals from color by simply using the
average globular colors with a small extrapolation to the reddest colors, and
use this value to compare with metallicity values determined from line indices.
However, as can be seen in all the panels of Figure \ref{color_feh}, the relationship
between [Fe/H] and color becomes very steep at the highest metallicities.  Small
errors in colors would magnify into very large errors in [Fe/H].  The 12 Gyr SSP
models (shown in red) greatly assist this extrapolation, as they follow the averaged
data very well and it only requires a small extrapolate to greater than solar
metallicities.  Thus, we can use the SSP models to guide the assignment of color to
globular [Fe/H] and, for galaxies, use a multi-metallicity model tied to the SSP 
and globular cluster [Fe/H] values.

Shown in Figure \ref{color_feh} are two SSP models (BC03) for 12 (red) and 5 (green)
Gyrs.  The 5 Gyrs model is 4 Gyrs younger than even the youngest globular, but is
shown for illustration.  As was seen in the two-color diagrams, there was frequently
a color offset, especially in the near-blue filters ($u$ and $g$).  All the SSP 12
Gyrs models were shifted in color to match the moving average values.  This
corresponds to a shift of $-$0.10 and $-$0.02 in $u-g$ and $g-r$, the other color SSP
tracks ($r-i$ and $g-K$) were in agreement with the mean globular colors.  We assume
this color offset is due to a mismatch between the conversion tables in the SSP
models to SDSS filters, and we simply make this correction in order to align the
measured [Fe/H] values to the mean globular colors in order to compare zeropoints.

Each globular color-metallicity diagram is well represented by the 12 Gyrs SSP
tracks.  There is no significant drop at high metallicities expected for slightly
younger clusters, although a color difference between a 9 and 12 Gyr population would
be difficult to detect given the photometric errors.  The top four color diagrams are
well-defined, but the $J-K$ metallicity diagram has a great deal of scatter.  This is
primarily due to the difference between M31 near-IR colors and the colors of MW
globulars from Cohen \etal (2015), shown as green symbols in Figure \ref{color_feh}).
The MW globulars are nearly 0.1 mags bluer for the same [Fe/H] than the M31
globulars.  The reason for this separation is unknown, but makes using $J-K$ colors
as a metallicity calibration problematic.

To test the accuracy of the globular cluster calibration, we use just the SSP models
to recover the globular cluster [Fe/H] values.  Each globular color is assigned a
separate [Fe/H] value, and the five colors are averaged an compared to the actual
[Fe/H] for each cluster.  The average of all five colors produces [Fe/H] value per
cluster with a dispersion of 0.2 dex, slightly better than the typical error of 0.3
based on index values.  Although this is simply a statement on how well the SSP
models match the [Fe/H] versus color relationships (Peacock \etal 2010).  The $J-K$
color has the worst predictive value, primarily due to the discrepancy between the MW
and M31 globular colors mentioned above.  The $g-K$ (i.e., $g-K$) color produced the
strongest match to globular [Fe/H] values.

To apply these models to ellipticals requires an expansion of the SSP models to the
type of composite stellar populations expected in galaxies.  Much of the following is
outlined in Schombert \& McGaugh (2014) and the reader is encouraged to review that
analysis for most of the details of the modeling.  In brief, the obvious next step
beyond a single age and metallicity SSP is to use single age SSP combined with a
range of metallicities that follows particular enrichment scenario (Pagel 1997).  Of
course, physically, it is impossible to generate enriched stars without a progression
in time of stellar birth and death, followed by recycling enriched material back to
the ISM.  This clearly takes a finite amount of time but, fortunately, the recycling
timescales are much shorter than noticeable changes in a stellar population due to
age (Matteucci 2007). Enrichment can be treated as instantaneous and the stars can be
assumed to arise from an enriched gas with a range of gas metallicities that reflect
the metallicity distribution function (MDF) of the stars of a single age.

For the current suite of models, we have followed the MDF prescription outlined in
Schombert \& Rakos (2009) that has the advantage of matching the chemical evolution
of Milky Way stars and correctly reproducing the narrow band colors of ellipticals.
The MDF model adopted in Schombert \& Rakos (2009, called the "push" model) is
designed to address the G-dwarf problem (Gibson \& Matteucci 1997, an underestimation
of low metallicity stars in closed box models) and incorporates a mixing infall
scenario (Kodama \& Arimoto 1997, linking the accretion rate to the star formation
rate).  While not as sophisticated as MDF deduced from various metallicity features
(see Tang \etal 2014), the colors extracted from a composite population are not very
sensitive to the exact shape of the MDF.  A range of MDFs are shown in Figure 1 of
Schombert \& McGaugh (2014), each model is parameterized by their peak [Fe/H].  At
low peak [Fe/H], the shape of the MDF is nearly gaussian and the average [Fe/H] for
the population is identical to the peak [Fe/H].  At higher metallicities, the shape
of the MDF is skewed to low [Fe/H] stars and the metallicity that can be assigned to
the stellar population has three states; 1) the peak [Fe/H], 2) the numeral mean
[Fe/H] from averaging all the bins or 3) the luminosity weighted [Fe/H], each bin
weighted by the luminosity of the stars in that bin (metal-poor stars are more
luminous than metal-rich stars).  Some numeral experiments demonstrated that the
relationship between peak [Fe/H] and the mean [Fe/H] (hereafter referred to as
$<$Fe/H$>$) was linear with peak [Fe/H] such that for super-solar values $<$Fe/H$>$
was $-$0.05 dex less than the peak.  The luminosity average was only relevant to
models where spectral line indices were observed and will be discussed in \S4.

Additional corrections due to the IMF, $\alpha$/Fe ratios, blue HB stars, blue
stragglers and TP-AGB populations are all discussed in Schombert \& McGaugh (2014),
and we have adopted these corrections to the multi-metallicity models where
appropriate for ellipticals (for example, ellipticals have a known increase in
$\alpha$/Fe with luminosity, Worthey \& Collobert 2003).  The resulting
multi-metallicity tracks are shown in Figure \ref{color_feh} as dashed lines for a 12
Gyrs mean age.  At low metallicities (values less than $-$0.7 [Fe/H]) the
multi-metallicity models are identical to the SSP models since the spread of [Fe/H]
narrows and approaches a gaussian in shape.  At higher metallicities, the effects of
the metal-poor component becomes apparent.  For a specific [Fe/H], the color of the
population is bluer, and the deviation is stronger in the bluer filters.  Thus, if we
wish to use the [Fe/H] values of globular clusters to convert colors into mean
[Fe/H], a correction must be made for the total color of the underlying stellar
population that includes a wider range of [Fe/H] stars.

Unsurprisingly, the multi-metallicity colors per [Fe/H] value are significantly bluer
than the SSP tracks (the so-called "red lean", Tang \etal 2014).  Using just SSP
tracks, one would deduce [Fe/H] values from a galaxy color that is much lower than
one deduced from the multi-metallicity tracks, and the discrepancy becomes larger for
bluer colors.  In other words, the [Fe/H] value one would deduce for a particular
ellipticals would be lower using $u-g$ than $g-K$ or $J-K$.  In addition, the
multi-metallicity tracks are shallower with [Fe/H], allowing for a larger range in
[Fe/H] for the range of elliptical colors.  One would conclude a much lower range in
[Fe/H] values using just SSP models.  The shallower slope for the multi-metallicity
models also has the advantage of lowering the error in [Fe/H] value due to
photometric error in color.

The multi-metallicity tracks, when tied to the globular cluster metallicities as a
zeropoint, provide a much more consistent picture of [Fe/H] with elliptical color.
Using the tracks, one can estimate an $<$Fe/H$>$ value, the mean metallicity for the
composite stellar population, for each galaxy's color.  Strictly from the five colors
displayed in Figure \ref{color_feh} ($u-g$, $g-r$, $r-i$, $g-K$ and $J-K$) one
obtains five values of $<$Fe/H$>$.  The scatter between the average in colors was
less than 0.2 dex in $<$Fe/H$>$ for all colors except for $J-K$ (probably due to the
poor globular calibration as seen in Figure \ref{color_feh}).  Dropping $J-K$ from
the average resulted in a scatter of 0.15 dex in $<$Fe/H$>$, which is similar to
dispersion in the globular sample.  

The resulting estimated $<$Fe/H$>$ values from the four colors are listed in Table 2
for five characteristic luminosities.  Note that these values are not a
substitute for [Fe/H] values deduced from line indices.  Those values measure an
actual metallicity feature.  The values in Table 2 simply reflect what mean
$<$Fe/H$>$ values would correspond to the five colors and, most importantly, what
range of [Fe/H] would be in conflict with the mean colors.  The $<$Fe/H$>$ values
for ellipticals, that correspond to the mean colors, range from $-$0.4 for the
faintest ellipticals in the sample to +0.4 for the brightest.  This is shown
graphically in Figure \ref{color_feh} as red symbols, the tight linear correlation
with [Fe/H] is, of course, an artifact of the fitting process, but demonstrates that
the same $<$Fe/H$>$ color is deduced from each of the five colors with a remarkable
degree of consistence given the various sources of uncertainty (the globular cluster
calibration, model track variation in stellar populations, photometric error in the
galaxy colors).

The $<$Fe/H$>$ values that match the observed colors is, however, in sharp
disagreement with canonical values extracted using Lick/IDS indices (see Graves \etal
2009; McDermid \etal 2015).  Most indices studies find much lower metallicity values
than found herein.  For example, Graves \etal (2009) finds that for $M_r = -22.6$
(equivalent to $M_J = -24.6$ in Table 2) the mean [Fe/H] value for an elliptical is
between 0.0 and $-$0.1 versus our expected value of +0.30.  At the faint end of their sample
($M_r = -19.4$, equivalent to $M_J = -21.3$ in Table 2), they deduce [Fe/H] values of
approximately $-$0.35 compared to our expected value of $-$0.02.  Age effects will not
reconcile the different metallicity determinations as younger ages will only drive
the mean colors bluer, which in turn will require even higher metallicities in the
multi-metallicity models to produce the observed colors in Table 2.  This
demonstrates the strong inconsistency between age and metallicity determined by studies
using spectral indices and the observed galaxy colors which we will return to in \S4.

Some of the disagreement between the Lick/IDS indices age and metallicity estimates
and our color method is due to the new multi-metallicity models.  For these models
predict bluer colors at any particular mean $<$Fe/H$>$ value simply due to the
inclusion of a metal-poor component.  However, even using the simpler SSP models
would produce discrepant colors, although to a lesser extent.  For example, at the
low luminosity end of our sample ($M_J = -21.3$), the predicted [Fe/H] value from
indices work is between $-$0.3 and $-$0.4 (see Figure 7, Graves \etal 2009).  That
metallicity value, when compared to a 12 Gyr SSP or globular cluster colors
corresponds to a $g-r$ color of 0.66, a $r-i$ color of 0.36 and $g-K$ color of 2.75.
These values are 0.06, 0.02 and 0.32 mags bluer than the mean colors in Table 2.  In
addition, at the low luminosity end of the elliptical sequence, Graves \etal deduce a
mean age of 6 Gyrs.  For a standard SSP, this would produce colors that were 0.16,
0.09 and 0.60 mags bluer than the mean colors in Table 2 (see also Schombert \& Rakos
2009).  The high end of the luminosity scale ($M_J = -24.6$, in Table 2) would have
ages near 12 Gyrs and metallicities near or slightly above solar from Graves \etal
(2009).  Again, from 12 Gyr SSP models, this would correspond to $g-r$, $r-i$ and
$g-K$ colors of 0.74, 0.39 and 3.45 compared to measured values of 0.83, 0.42 and
3.87, i.e. consistently too blue from observed values.  These differences would even
be more dramatic if the multi-metallicity models were used as they have even bluer
model colors per [Fe/H] than the SSP models.

Lastly, it needs to be considered whether the blueward trend for elliptical colors,
compared to globular SSPs, is due to an age effect.  Under the standard scenario
proposed by line indices studies, the ages of brighter, redder ellipticals is less
than the mean age of the stellar populations in faint ellipticals.  This age
difference can reflect different initial star formation epochs ($\tau_{iSF}$) or be
the result of a longer duration of initial star formation ($\Delta\tau_{iSF}$) where,
in this circumstance, the more massive ellipticals have longer durations of initial
star formation resulting in a larger fraction of younger stars (see Thomas \etal
2005).  Longer star formation durations would seem to be ruled out by $\alpha$/Fe
ratio studies of ellipticals (Worthey \& Collobert 2003) that display higher
abundance of $\alpha$ elements with redder galaxy color indicating shorter initial
star formation durations (less time for SN type Ia's to input their Fe-rich component
to the galactic ISM).  Younger mean initial formation age would shift the
multi-metallicity tracks towards the green 5 Gyrs track in Figure \ref{color_feh} and
slightly younger ages (between 8 and 10 Gyrs) could match the elliptical colors if
higher mean metallicities were used.  

As noted in the last section, the two-color diagrams are useless for testing if the
conclusions from line indices studies, concerning a decrease in mean age with
elliptical luminosity, is true.  However, each color predicts a unique [Fe/H] which,
when calibrated with 12 Gyrs multi-metallicity tracks, produces a set of predicted
$<$Fe/H$>$ values that have the same mean value within the errors.  While these
values are not superior to actual [Fe/H] measurements using indices such as [MgFe] or
$<$Fe$>$, they can bee used as a guide to the comparison between continuum determined
values, such as colors, and spectral indices.  To this end, we can allow the model
mean ages to vary and test the predicted $<$Fe/H$>$ values for each color with these
new tracks.  Using 5 Gyr tracks (the youngest ellipticals from Graves \etal and
McDermid \etal are between 5 and 6 Gyrs), we find that the lowest luminosity
ellipticals have mean metallicities from blue colors (e.g., $u-g$) to be around
solar.  But, for the near-IR colors, such as $g-K$, the predicted metallicity is
+0.5.  In other words, lower mean ages produce conflicting metallicities values
deduced from their colors (likewise deviant colors for the metallicity).  Ages
between 10 and 13 Gyrs produced consistent $<$Fe/H$>$ values from color to color, but
ages below 9 Gyrs resulted in conflicting colors at all luminosities and predict
extremely large $<$Fe/H$>$ values in excess of metallicities expected from line
indices using Mg$b$ or [MgFe] (see the next section).

\section{Colors and the Lick/IDS Indices}

Improved technology, and larger telescope apertures, led to the replacement of colors
as indicators of stellar populations with more high resolution spectral indices
(Trager \etal 2000). With the addition of improved SED models, it was found that
spectral indices alone could deduce the age and metallicity of underlying stellar
populations (Kuntschner 2000; Trager \etal 2000).  The method of determining age and
metallicity of a galaxy matured with the introduction of the Lick/IDS line-strength
system that, primarily, depends on Fe lines (notably Fe5270 and Fe5335), Mg$_b$, and
H$\beta$ lines to deduce mean age and metallicity (Trager \etal 2000). A surprising
result from the spectroscopic surveys (e.g., Gallazzi \etal 2006) is that a
significant fraction of early-type galaxies have mean ages younger than expected from
monolithic scenarios (see Schiavon 2007 for a review).  The technique reached its
peak with the use of the SDSS dataset to study the age and metallicities of over
16,000 galaxies in the red sequence (Graves \etal 2009, 2010).

\begin{figure}[!ht]
\centering
\includegraphics[scale=0.75,angle=0]{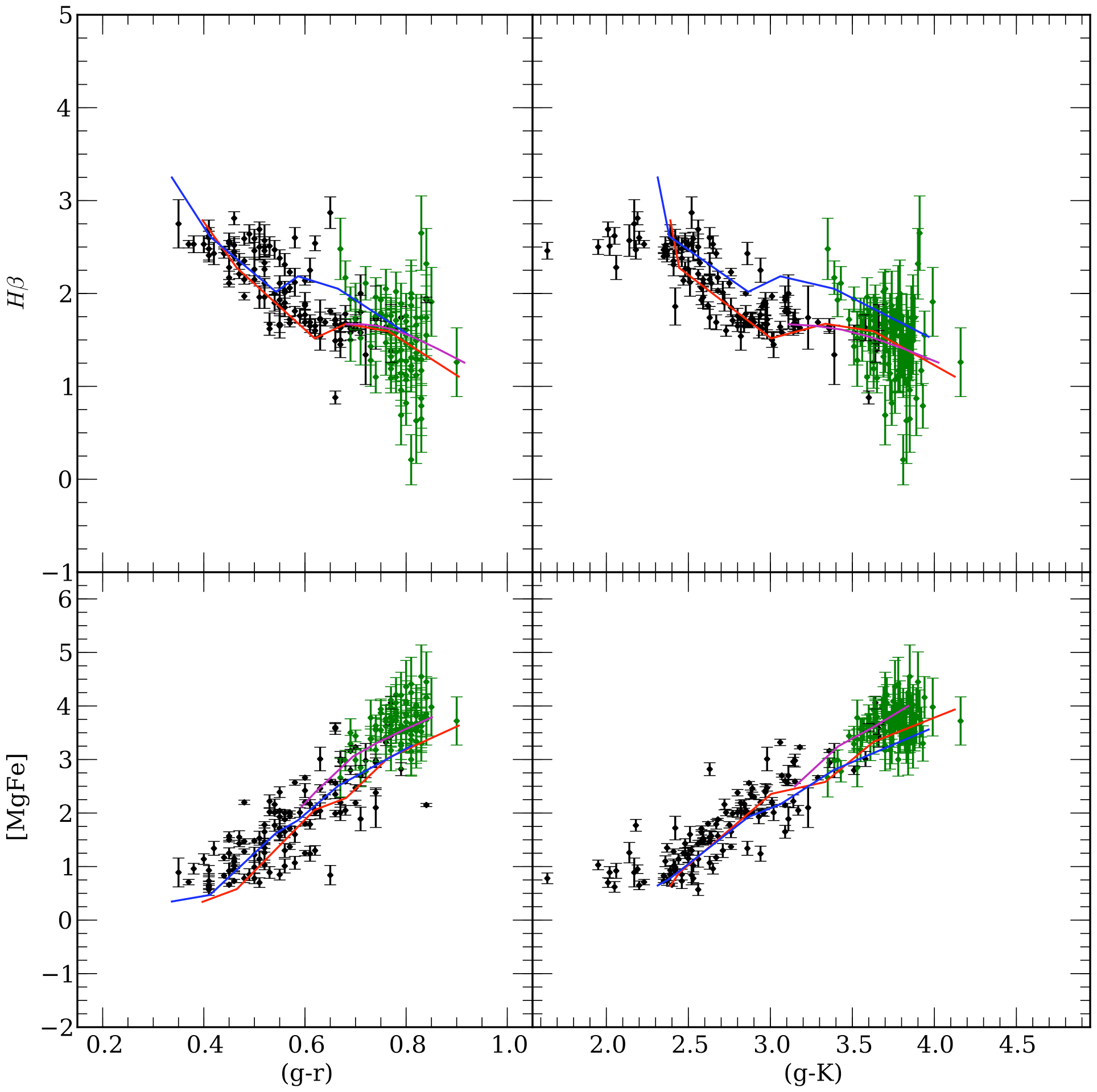}
\caption{\small The color-Lick/IDS indices diagrams for [MgFe] (the metallicity index)
and H$\beta$ (the age index).  The black symbols are the globular clusters with
available spectral data.  The red symbols are 91 ellipticals in common between our
sample and spectral studies.  The red and blue tracks are  and 12 Gyrs SSP tracks.
The magenta line is a 12 Gyr multi-metallicity track.
}
\label{lick_avg}
\end{figure}

For comparison to our sample, we have extract 91 ellipticals with Lick/IDS indices values
from Trager /etal (1998).  The Lick/IDS indices [MgFe] and H$\beta$ versus $g-r$ and
$g-K$ are shown in Figure \ref{lick_avg} along with the globular data, SSP and
multi-metallicity models from the previous sections (other colors display the same
behavior as these two colors).  For globulars, we have taken the colors from Galleti
\etal (2009), using the same $E(B-V)$ and apparent magnitude cutoff as \S3.4, and the
indices values from Schiavon \etal (2012).  The [MgFe] index is closely coupled to
[Fe/H], and minimizes the effects from changes in $\alpha$/Fe (Galleti \etal 2009).
The H$\beta$ index is closely coupled to mean age or recent star formation (Trager
\etal 1998).  For globulars and ellipticals, this is expected to reflect mean age as
there is no evidence of recent star formation from the $NUV$ colors.  As a caveat,
the spectroscopic values are derived from small aperture observations, which will be
compared to large aperture colors.  Nuclear starbursts are not uncommon in
ellipticals, which would skew the H$\beta$ values.  However, there are no signatures
of recent star formation in the $NUV$ colors (particularly at small radii) and strong
emission lines are usually associated with active nuclei, which were excluded from
the Trager sample.

The trend between globulars and ellipticals with colors continues for the Lick/IDS
indices.  Both globulars and ellipticals display a steady increase in [MgFe], and a
decrease in H$\beta$, with redder colors.  Also shown in Figure \ref{lick_avg} are
two population tracks from BC03 for 12 and 5 Gyrs SSPs, and multi-metallicity tracks
for a 12 Gyrs composite population.  As can be seen in the lower panels, [MgFe] is
relatively insensitive to age, and both models follow the globulars data to within
the spectroscopic and photometric errors.  The H$\beta$ index is a clearer indicator
of age and the globulars are well-matched to the 12 Gyrs SSPs, although the observed
H$\beta$ values are slightly higher on average.  Newer Lick/IDS indices tracks (e.g.,
Schiavon 2007) match the globular data better (see Figure 9 in Caldwell \etal 2011)
and lack the "bump" seen at $g-r = 0.6$.  This is probably due to a better treatment
of BHB stars and other hot, metal poor components, but both set of tracks display
the same endpoints in index value versus color, and either models produce the same
results.

The ellipticals continue the globular trend with [MgFe], but lie too high in [MgFe]
(or too blue in color) compared to the SSP models.  Changing age will not reach these
portions of the [MgFe]-color diagrams; however, this is the expected deviation for a
composite population.  The multi-metallicity tracks in Figure \ref{lick_avg} recover
the observed [MgFe]-color relation for ellipticals and we note that, simply through
comparison with the color versus [MgFe] values of globulars, very few ellipticals
have [Fe/H] values less than the most metal-rich globulars (nearly solar).  Lower
mean ages will not resolve this issue as they will only increase the deduced [Fe/H]
from [MgFe].  

A composite multi-metallicity model is even more critical in the near-IR colors.  The
[MgFe] values for ellipticals are much higher than the globular values in the $g-K$
diagram.  However, the colors are well described by the 12 Gyrs composite track
(magenta line in Figure \ref{lick_avg}).  As we have seen with the two color
diagrams, different age models have the same color-[MgFe] slope, but differ on what
value of [Fe/H] would be deduced from a single [MgFe] value.  Again, the fact that the
elliptical data differs from the globular index values versus color slope argues for
a composite population in ellipticals.

The behavior of the H$\beta$ index for ellipticals is a more significant measure of
mean age.  The H$\beta$ index, when plotted against [MgFe], is the primary source of
all the claims for ellipticals with ages of less than 10 Gyr and rests at the
heart of most hierarchical merger galaxy formation scenarios.  The mean H$\beta$
values in Figure \ref{lick_avg} are slightly higher than the 12 Gyrs SSP and the 12
Gyrs multi-metallicity model predicts higher H$\beta$ values than the 12 Gyrs SSP due
to the contribution from metal-poor stars.  In fact, the multi-metallicity 12 Gyrs
model divides the elliptical data evenly.  There is a weak luminosity-H$\beta$
correlation, the ellipticals with high luminosities have slightly lower H$\beta$
values and vice-versa for low luminosity ellipticals.  This relationship is often
used as evidence that elliptical ages vary with stellar mass (younger for lower mass
ellipticals).  However, the 12 Gyr composite track decreases in H$\beta$ with
increasing [Fe/H] and redder color.  This would reflect into lower H$\beta$ values
for high mass ellipticals due to the mass-metallicity relation.  In other words,
there is no evidence from the color-H$\beta$ diagram for ages younger than 12 Gyrs
for any ellipticals.  One might be inclined to deduce younger ages based on the
optical colors.  However, the near-IR colors display different behavior.  In the
$g-K$ diagram, the H$\beta$ values are well aligned with the 12 Gyrs
multi-metallicity tracks and the average value lies significantly below the 5 Gyr
tracks.  This trend is evident in all the near-IR colors.

In addition, for the highest luminosity ellipticals, their Lick/IDS indices predict
near solar metallicities at a $g-r$ color of 0.80.  At low luminosities, the color
drops to 0.72 with metallicities corresponding to [Fe/H] = $-$0.3 to $-$0.4.
However, just through comparison of globular cluster colors, a globular at $g-r =
0.72$ has a [Fe/H] value of $-$0.2 and for $g-r = 0.80$ we can extrapolate the
globular metallicity to +0.3.  This is much more metal-rich than the conclusions from
Graves \etal (2009) that find the brightest ellipticals to only have slightly higher
than solar metallicities.  We conclude that 1) either the Graves \etal colors are
incorrect or 2) the metallicity values they propose are too low.  However, even if
their high luminosity colors are too blue, correcting them to our value of 0.83
produces expected [Fe/H] values near +0.5, which is ruled out by their own [MgFe]
values.  Younger ages for the reddest galaxies would only decrease the $g-r$ colors.
Whereas, their own analysis calculates a mean age for the reddest galaxy of 12 Gyrs,
similar to globulars.  The disconnect between color and age/metallicity, as deduced
from Lick/IDS indices, is evident whether our colors or theirs are used for
comparison.

\begin{figure}[!ht]
\centering
\includegraphics[scale=0.75,angle=0]{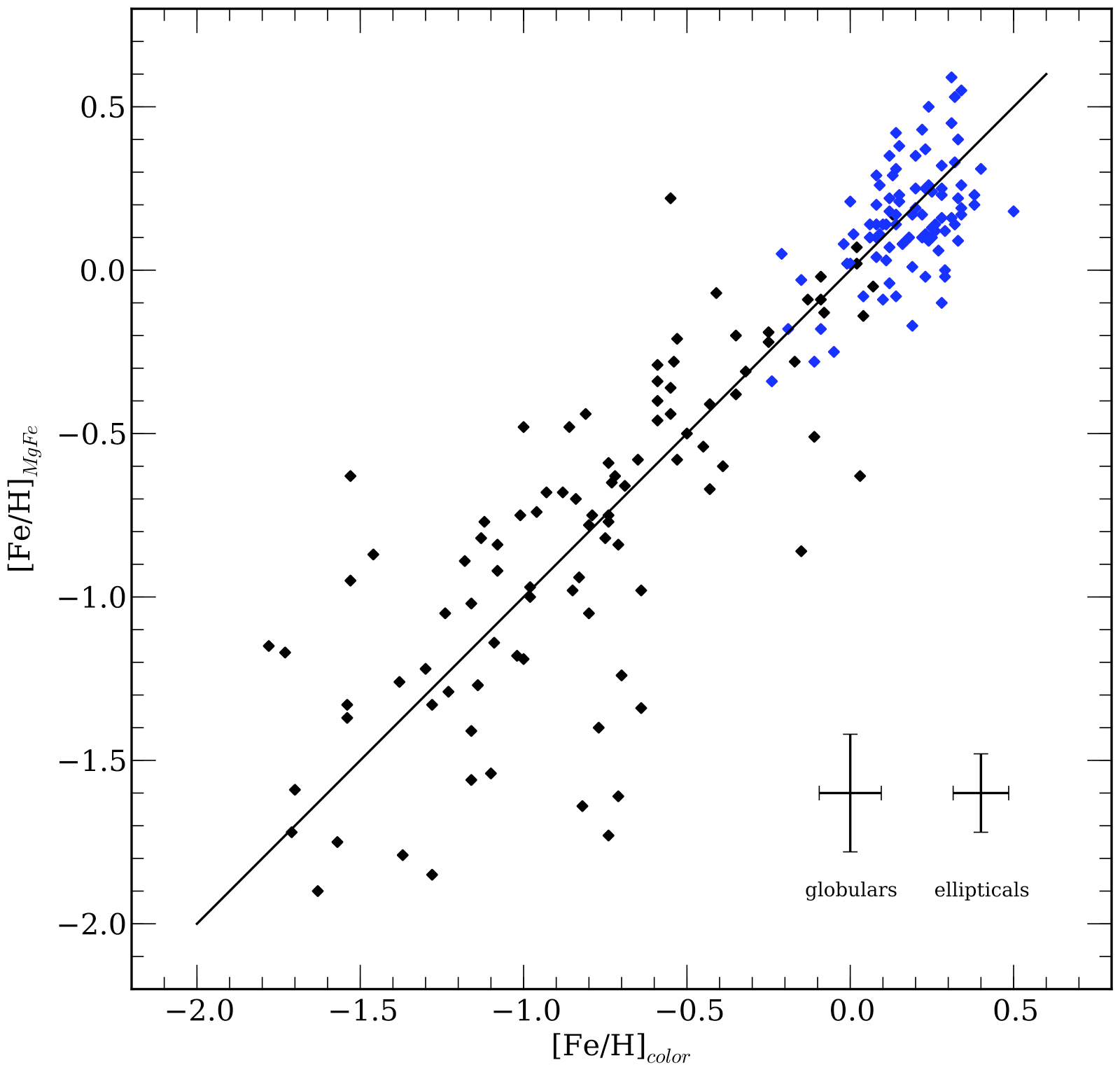}
\caption{\small Comparison of [Fe/H] deduced from the Lick/IDS [MgFe] index and colors.
The [MgFe] are converted to [Fe/H] using the prescription of Galleti \etal (2009) for
globulars (black symbols) and the composite population models for ellipticals (blue symbols).
The [Fe/H]$_{color}$ values are derived from the relations defined in \S3.4 using five
colors from $u$ to $K$.  The dispersion from the unity line for both globulars and
ellipticals is consistent with the uncertainties in the observations.  The use of
colors to determine [Fe/H] matches the accuracy for spectroscopic observations.
}
\label{feh_color_mgfe}
\end{figure}

Lastly, we can use the [MgFe] index as a check to the [Fe/H] values expected
from the optical and near-IR colors.  This comparison is shown in Figure
\ref{feh_color_mgfe}.  For the globulars, the [Fe/H]$_{MgFe}$ values are, in fact,
the [Fe/H] values quoted in Galleti \etal (2009) using a polynomial fit to MW
globulars.  The [Fe/H]$_{color}$ values are from the SDSS and near-IR photometry
quoted in \S3.4.  The ellipticals [Fe/H] values are the ones deduced from the
composite population models in \S3.3, with the same models applied to the [MgFe]
index.  Average error bars are shown.  Most of the error in the globulars is from
spectroscopic errors in the [MgFe], although the photometric errors are not
negligible.  The errors on [Fe/H]$_{color}$ for ellipticals derives from photometric
errors (which are small) and the dispersion surrounding the [Fe/H] value derived from
the average of five colors ($t-g$, $g-r$, $r-i$, $g-K$, $J-K$).  The correspondence
is good, the dispersion in [Fe/H]$_{color}$ for globulars matches the uncertainty in
the deduced [Fe/H] values from the [MgFe], which were used to set the zeropoint of
the color-[Fe/H] relations in \S3.4.  The dispersion for ellipticals is also similar
to the uncertainties from the color-[Fe/H] relations, but a key point is that the use
of five colors produces similar [Fe/H] values as the, much more difficult to obtain,
spectroscopic values.

We note that the line indices values are core measurements and, as can be seen
in Figure \ref{frac_colors}, the core colors are slightly redder than half light
colors due to color gradients.  This effect is fairly minor (only 0.03 in $g-r$) and
works in the opposite direction needed to reconcile colors with the age and
metallicity values deduced from line indices.  The comparison in Figure
\ref{feh_color_mgfe} reinforces the disparity between colors and line indices values, the
use of colors is by no means a superior technique to using line indices to determine
age and metallicity.  Colors measure the continuum flux over a range of wavelengths.
The continuum varies with temperature of the underlying stars which is an {\it
indirect} measure of age and metallicity.  Metallicity indices, such as [MgFe] are
more direct measures of the quantity of interest, the [Fe/H] value for the majority
of the stars.  It is easy to understand how small errors in the calibration of
stellar libraries could lead to erroneous colors, yet produce accurate indices values
due to the smaller range in wavelength coverage.

\section{Color-Magnitude Diagrams}

One of the earliest fundamental relationships for ellipticals is the color-magnitude
relation (CMR).  Originally based on photoelectric color measurements (Sandage \&
Visvanathan 1978), the CMR reflects a correlation between the mean color of an
early-type galaxy and it's total luminosity (luminosity assumed to be a proxy for
stellar mass).  While other scaling relations for ellipticals (e.g., the
Faber-Jackson relation, Faber \& Jackson 1976, and the Kormendy relation, Kormendy
1985) outline, primarily, the kinematic and structural correlations related to
formation processes, the CMR is driven by stellar population effects as indicated by
the closely related color-$\sigma$ and color-size relations.  Observations of
ellipticals over a range of environments (cluster to field, Bower \etal 1992) and to
high redshift (Yabe \etal 2014) demonstrates the universality of the CMR.  In
addition, the CMR has been investigated from the far-UV (Jeong \etal 2009) to the
far-IR (Clemens \etal 2011), plus extended from giant to dwarf ellipticals (Rakos \&
Schombert 2005).  In order to follow the evolution of the CMR with redshift, one must
define the present-day CMR to a high degree of accuracy and in bandpasses that will
be observed at high redshift (i.e., the blue and UV).  

The explanation for the CMR has focused on two processes.  The first, and probably
the most dominant, is an increase in mean metallicity of the underlying stellar
population with increasing stellar mass (Schombert \& Rakos 2009).  The
mechanism herein is simply that larger galaxies have deeper gravitational wells and
retain more metal-enriched stellar ejecta for each generation (Matteucci 2007).  This
raises the mean metallicity of the stellar population resulting in redder optical and
near-IR colors.  An increase in the mean [Fe/H] with galaxy mass has been supported
by $<$Fe$>$, Mg$_b$ and other metallicity indices values (Graves \etal 2009; McDermid
\etal 2015).

Age or recent star formation can also factor into the slope of the CMR, for age
effects serve to push optical colors blueward (whether a small young stellar
component to the elliptical or a later epoch of initial star formation, see Faber
\etal 1995).  Both metallicity and age are difficult to distinguish in colors (due to
the age-metallicity degeneracy, see Worthey 1994), although lower age works in the
opposite direction to the metallicity clock (Thomas \etal 2005).  The two effects can
conspire together if metallicity increases with galaxy mass, as expected from
chemical evolution models, and the mean stellar age (or recent star formation) is
more dominate in lower mass ellipticals.  Mergers can also play a role, especially if
gas-rich disks are the progenitors.  This process would play havoc with the
integrated colors of low luminosity ellipticals (randomly undergoing mergers of
nearly equal mass) versus brighter ellipticals that would, statistically, only
absorb a small percentage of their mass by mergers.

\begin{figure}[!ht]
\centering
\includegraphics[scale=0.75,angle=0]{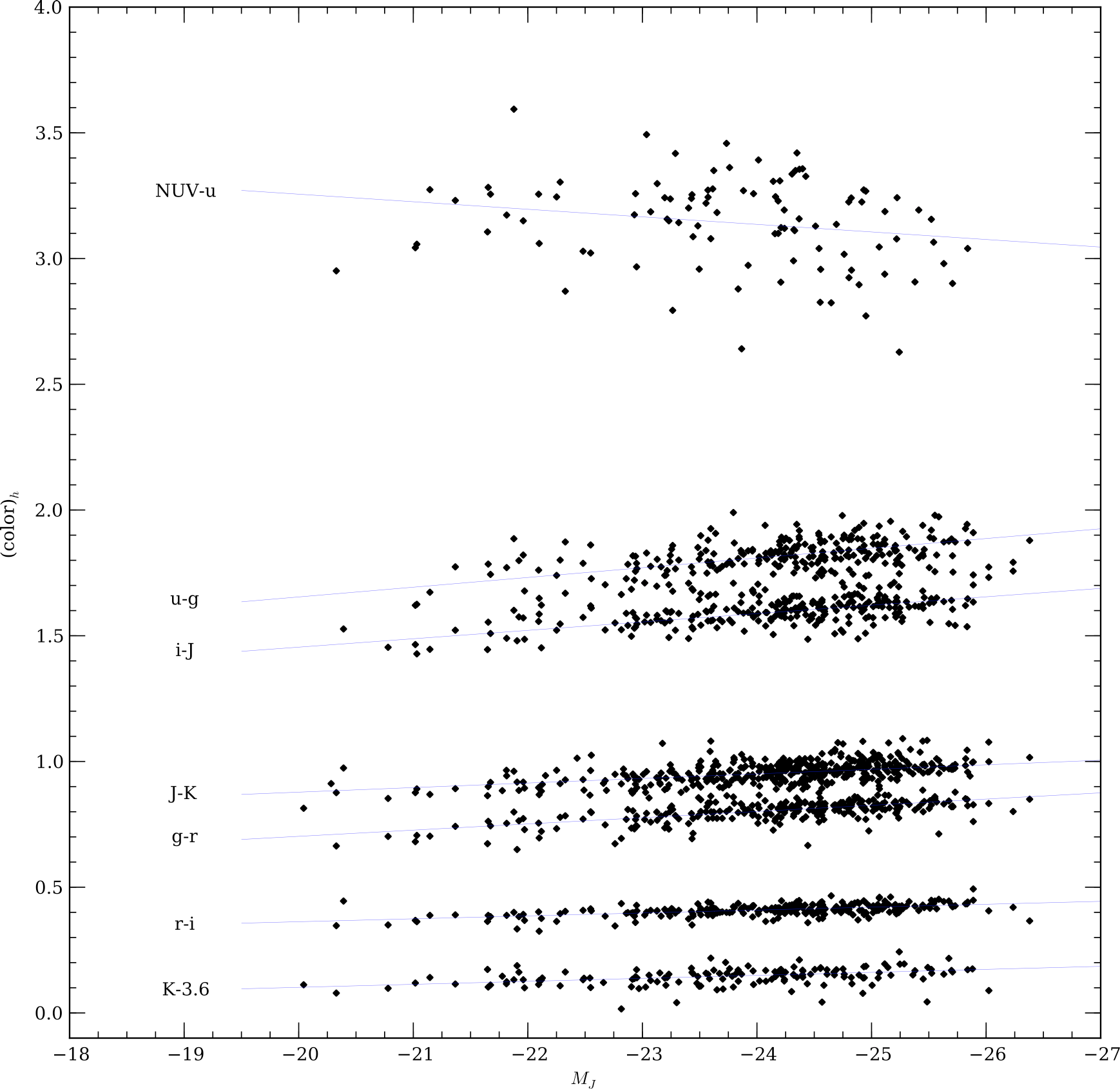}
\caption{\small The color-magnitude relation for seven of the colors in our sample.
The trend of increasing slope with decreasing wavelength is evident, with the
expectation of the $NUV$ colors.  The slopes compare favorably with other studies in
the optical and near-IR (Bernardi \etal 2003; Chang \etal 2006).  The dispersions
around each CMR are identical to the dispersions for the mean colors found in Table 2
indicating the scatter is purely photometric.
}
\label{cmr_all}
\end{figure}

For our sample, we have used the total luminosity at $J$ for the magnitude axis and
the half-light radius color for the color axis.  Each of the colors for this study
were fit by a jackknife linear algorithm with the resulting slopes standard
deviations tabulated in Table 3.  The fits for eight filters, and the resulting seven
neighbor colors, are shown in Figure \ref{cmr_all}.  The dispersions around each CMR
matches the dispersion in color as shown in Table 2, which were demonstrated to be
purely photometric in nature.  There is no indication of non-linearity to the
relationship (over our limited luminosity range) and residuals to the CMR were
uncorrelated with any galaxy parameter (e.g., scale length, effective surface
brightness, etc.).  

All the slopes are identical, to within the fitting and photometric errors, to the
CMR slopes from many other studies.  For example, Bernardi \etal (2003) and
Chang \etal (2006) both find a slope of $-$0.025 for the $g-r$ CMR, which is
identical to our slope in Table 3.  In addition, Chang \etal finds slopes of
$-$0.0115 for $r-i$ and $-$0.0182 for $J-K$, compared to our values of $-$0.0116 and
$-$0.0181, the agreement is remarkable.  Over longer wavelength baselines, the
agreement is less consistent.  We find a CMR slope for $g-K$ of $-$0.0878 versus the
value of $-$0.0917 from Chang \etal, a slight difference within the fitting errors.

There are several notable trends in Figure \ref{cmr_all}.  First, aside from the
$NUV-u$ CMR, all the CMR's have negative slopes consistent with the average colors at
each luminosity.  Second, there is no indication on non-linearity in any of the
colors.  The residuals from a linear fit were uncorrelated with any photometric or
structural property of ellipticals.  Lastly, the colors at the high and low
luminosity ends are consistent with the colors in each CMR and reproduce $<$Fe/H$>$
values (from the relationships defined in \S3.4) to the same metallicity value,
within the errors.  

As with the two-color diagrams, it is difficult to discern any differences between
metallicity and age effects simply from a single CMR.  However, the slopes of the all
the CMRs in Table 3 can be completely explained solely by metallicity effects without
any need for younger populations.  However, for each individual color, there is
always some combination of age and metallicity that can reproduce a given color.  For
example, for an 0.5 $L_*$ elliptical, the mean $g-r$ color is 0.780 (see Table 4).
That color is obtained in a 12 Gyr population for a [Fe/H] value of +0.20, but, if
one uses a 6 Gyrs population, the same color can be obtained from a [Fe/H]=+0.45
metallicity.  Although to achieve these colors, one must adopt a metallicity value
that is directly at odds with indices studies which indicate lower metallicities for
lower mass ellipticals (Graves \etal 2009; McDermid \etal 2015).

To explore this more fully, we consider younger ages for low and intermediate
luminosity ellipticals (it is universally agreed upon by indices studies that high
mass ellipticals are old and metal-rich).  In Table 4, we tabulate the effects of
younger age on elliptical colors from $u-g$ to $J-K$ for a $M_J = -$21.3 and $-$23.1
elliptical (0.1 and 0.5 $L_*$).  Indice studies deduce a mean age of 6 Gyrs for low
luminosity ellipticals.  The colors for a composite 6 Gyrs, [Fe/H]=$-$0.02 is shown
in Table 4 and display colors that are much too blue compared to the average values
in Table 2.  If we assume the metallicity value is incorrect, and raise the [Fe/H]
value such that the model $g-r$ color matches the observed $g-r$ color, we obtain the
model shown in the third line in Table 4.  While this model, which requires an [Fe/H]
of +0.30, correctly matches the $g-r$ and $r-i$ colors.  But, predicts a bluer $u-g$
color and redder $g-K$ and $J-K$ colors.  A similar experiment on a $M_J = -23.1$
galaxy arrives at a similar dilemma.  A 9 Gyrs composite population of mean solar
metallicity is too blue across all the colors.  Raising the metallicity to +0.31
matches the $g-r$ and $r-i$ colors, but again, predicts a bluer $u-g$ color and
redder $g-K$ and $J-K$ colors.  The range in acceptable ages to match all the colors
from $u$ to $K$ is vary narrow and can not exceed 2 Gyrs, older or younger than 12
Gyrs, for the range of stellar mass explored in our sample without producing mean
colors outside the acceptable color dispersions per luminosity bin.  In other words,
as first demonstrated by Kodama \& Arimoto (1997) and sequentially by Schombert \&
Rakos (2009), {\it it is extremely difficult to achieve the observed CMR of ellipticals with
mean ages less than 10 Gyrs, or extreme metallicities, at any luminosity}.

The CMR also maps into the color-$\sigma$ relation, where the central velocity
dispersion is used instead of total luminosity.  The correlation between luminosity
and velocity dispersion (the Faber-Jackson relation) guarantees this connection.  For
a limited number of ellipticals in our sample, we have velocity dispersion
information and the relationship is the same as presented in Bernardi \etal (2003)
and Chang \etal (2006).  We note that the scatter between color and velocity
dispersion is similar to the CMR's even though the accuracy of the central velocity
dispersion values is less than the photometric errors on luminosity, and we are
mapping a global characteristic of an elliptical (color) versus a core value
(velocity dispersion).

\section{Near-UV Colors and the UV Upturn in Ellipticals}

The behavior of the $NUV$ colors displays behavior seen before in ellipticals.  The
so-called "UV upturn" problem has been known since IUE and HUT (Bertola \etal 1982,
Brown \etal 1997).  The UV upturn is defined as a increase in luminosity for
early-type galaxies, between the Lyman limit and approximately 2500\AA, above what is
expected for a old stellar population.  An increase in this region is normal for an
object with recent star formation due to the contribution of massive OB stars.  However,
ellipticals are typically devoid of bright OB stars, thus the origin of the UV upturn has
remained uncertain.

While originally thought to be a phenomenon that is common to all bright ellipticals
(Yi \& Yoon 2004), other studies have found only a small fraction of ellipticals to
display a strong shift in flux in the UV (Yi \etal 2005).  The {\it GALEX NUV} band
is slightly redward of the canonical UV upturn region, but the reversed two-color
behavior in Figure \ref{6color_split} for $NUV-u$ indicates the $NUV$ colors are
unusual compared to the optical and near-IR colors.  Donas \etal (2007) find a
positive correlation between $NUV-V$ versus $B-V$, which would be equivalent to our
$NUV-g$ versus $u-g$.  We also find a weak, positive slope correlation between
$NUV-g$ versus $u-g$; however, any comparison between $NUV-u$ and any color displays
a inverse correlation with decreasing $NUV-u$ colors for redder galaxies (i.e.,
higher mass and higher metallicities).  The data from Smith \etal (2012), a study of
150 ellipticals in Coma, also finds a reverse slope in their $NUV-u$ CMR, although
they failed to notice this in their analysis.

\begin{figure}[!ht]
\centering
\includegraphics[scale=0.75,angle=0]{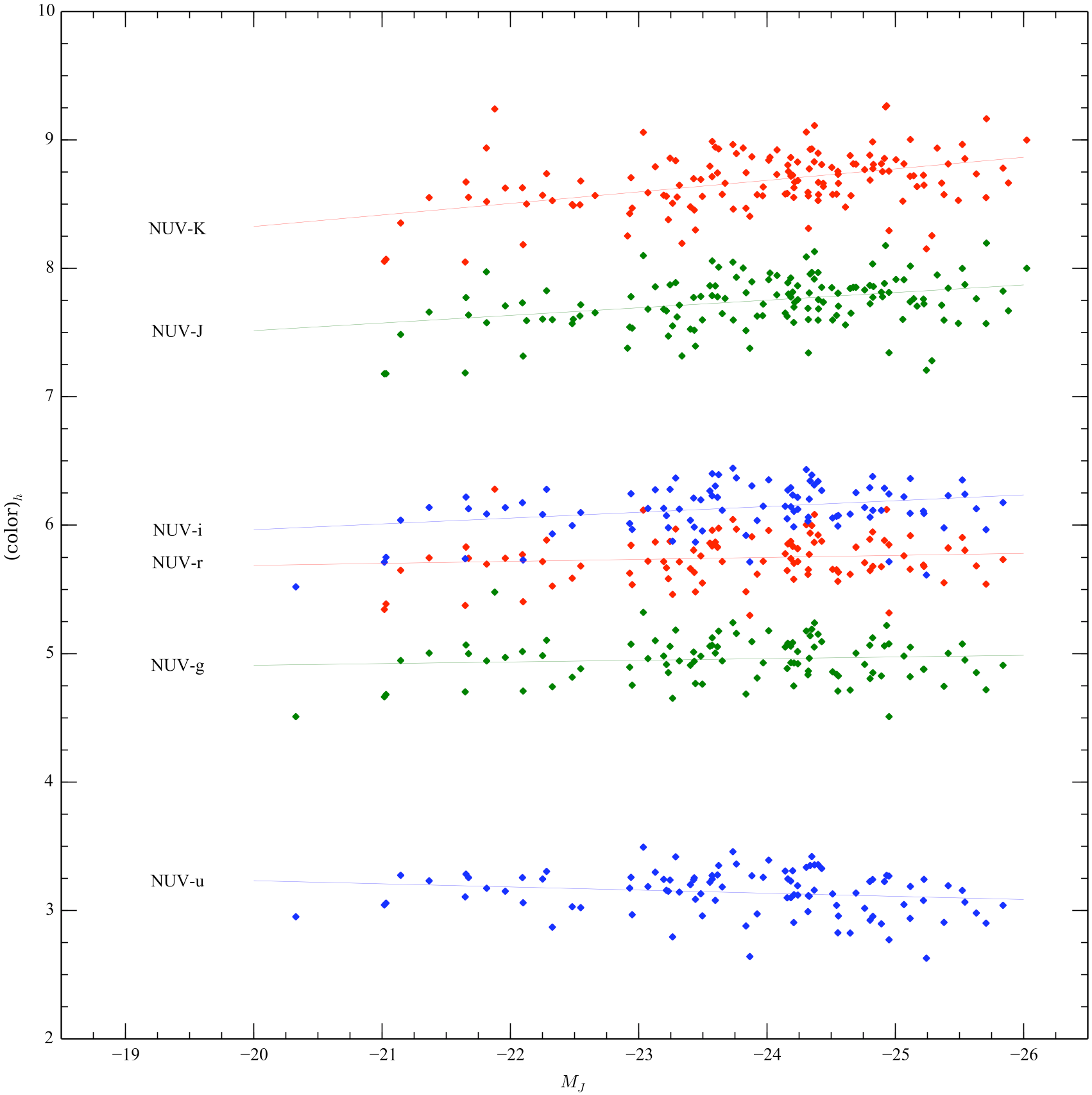}
\caption{\small The $NUV$ CMR with respect to all the filters in our study.  The
trend to go from positive slopes back to the negative slopes found in optical and
near-IR colors is evident.  Only the colors closest in wavelength to the $NUV$
bandpass display the unusual behaviour described as the "UV upturn" in the
literature.
}
\label{cmr_galex}
\end{figure}

The $NUV$ CMR also deviates from the normal behavior of optical and near-IR CMR's by
having a downward slope with respect to luminosity (see Figure \ref{cmr_galex}) for
the $NUV-u$ color with increasingly positive slopes for redder colors (nearly flat
for $NUV-g$, highest for $NUV-K$).  This behavior was seen in Jeong \etal (2009),
although they lacked $U$ data to see the inverse behavior for filters close to
2800\AA\ .  The scatter in the various $NUV$ CMRs in Figure \ref{cmr_galex} is
primarily due to error in the $NUV$ photometry and the higher Galactic extinction
corrections to the near-UV.  Like the CMRs from other colors, there is no evidence of
non-linearity to the relationships.

\begin{figure}[!ht]
\centering
\includegraphics[scale=0.75,angle=0]{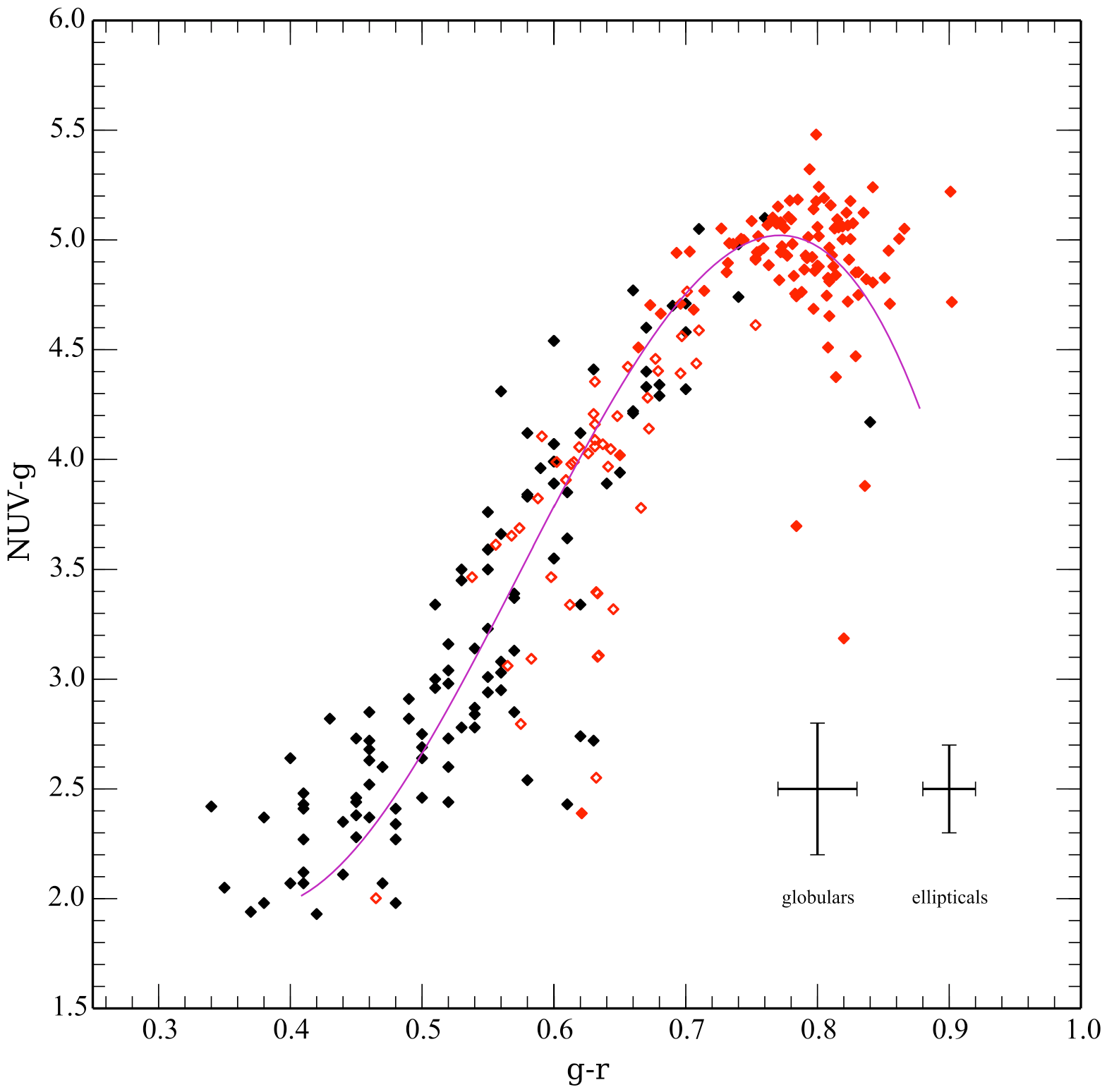}
\caption{\small The clearest signature of unusual UV color behavior is seen in
$NUV-u$ and $NUV-g$.  The $NUV-g$ versus $g-r$ colors of globulars (black symbols),
normal ellipticals (solid red symbols) and dwarf ellipticals (open red symbols) is
shown.  The normal increase in color for the globulars is evident; however, the
$NUV-g$ color behavior reverses near $g-r = 0.75$ where low luminosity ellipticals
increase in $NUV-g$ color and high luminosity ellipticals display decreasing $NUV-g$
color with $g-r$ color.  The model track displays BHB models of Yi \etal (1998) where
we have converted their [Fe/H] values (from -1.7 to 0.3) into $g-r$ color.  The
agreement with the data is excellent, even predicting the change in $NUV-g$ color
behavior with elliptical color.
}
\label{nuv_model}
\end{figure}

A clearer picture of the behavior of the $NUV$ colors can be seen in Figure
\ref{nuv_model}.  Here we have plotted the $g-r$ versus $NUV-g$ color for globulars
and ellipticals (a sample of dwarf ellipticals is added taken from a study of dwarf
colors, Schombert 2017).  The globulars form a well-defined sequence where it appears the
$NUV$ colors are driven strictly by metallicity.  The ellipticals form an interesting
sequence with brighter ellipticals decreasing in $NUV-g$ color with higher
luminosity.  The $NUV-g$ color is maximal around $g-r = 0.75$, then begins to drop
again to match the globular sequence.  This behavior in the $NUV-g$ clarifies
statements in the literature such that only bright ellipticals in clusters display
the UV upturn (Yi \etal 2005).  It appears the UV upturn is a smooth, non-linear
change in color where ellipticals transition from the globular sequence of redder
$NUV-g$ color to bright ellipticals that display an increasing UV upturn component
and, thus, bluer $NUV-g$ colors.  

Interpretation of the $NUV$ colors is difficult.  Standard SSP models predict a
change in $NUV$ flux that is solely due to recent star formation (SF).  Recent SF is
expected to be very minor in our sample due to the $NUV-r$ color selection and the
lack of any sharp discontinuities in color that would signal HII regions or strong
emission lines.  This is in concordance with the study done by Rutkowski \etal
(2013), which found that for ellipticals between redshifts of 0.35 and 1.5 less than
10\% of their stellar mass involved in a recent SF event ($\tau < 1$ Gyr).  As the
low redshift end of this sample corresponds to ellipticals with lookback times of
less that 2 Gyrs ago, even those minor SF events would have faded from detectable at
the current epoch.

Ruling out massive OB stars as the source of $NUV$ flux, one naturally turns to other
hot star components in old populations.  Leading contenders are blue horizontal
branch (BHB) stars, blue stragglers and planetary nebula cores.  The apparent smooth
transition from low luminosity to high luminosity ellipticals $NUV-g$ color suggests
a metallicity effect plays an important role, as this is the only underlying
population variable to change over this luminosity range.  This scenario is also
supported by the correlation between $Mg_2$ index and the UV upturn (Donas \etal
2007).  Blue stragglers relate to the evolution of the binary population and become
constant after 5 Gyrs with no change due to metallicity.  Planetary nebula cores can
vary slightly with metallicity, but are also short-lived and very low in luminosity.

The strongest candidate for the UV upturn is the BHB population for metal-poor BHB
stars are hot and contribute to the portion of the UV spectrum observed by the $NUV$
filter.  In addition, there are studies that suggest that metal-rich HB stars also
end in a UV bright phase (Horch \etal 1992).  The trigger for the UV upturn, with
respect to the underlying stellar population, may be a combination of increasing
metallicity and increasing $\alpha$/Fe ratios (Yi 2008).  

To examine this hypothesis, we have taken the single abundance models from Yi \etal
(1998) that uses an empirical mass-loss formula to reproduce $m_{2500}-V$ colors as
a function of [Fe/H].  We have converted $m_{2500}-V$ to $NUV-g$ and tied the model
[Fe/H] values to $g-r$.  The resulting track is shown in Figure \ref{nuv_model}.  The
Yi \etal scenario seems surprisingly well-matched to the behavior of the $NUV-g$
colors.  Surprisingly in the sense that the number of assumptions and exotic elements
of stellar evolution that go into these simulations make extracting precise tests to
the models problematic.  But, the models do predict the correct change in $NUV$ colors
with galaxy luminosity certainly this avenue of investigation seems promising with
higher quality far-UV data on ellipticals, particularly at lower luminosities.

\section{Discussion and Conclusions}

The Lick/IDS indices are often used to deduce age and metallicity, typically, through the
H$\beta$ versus [MgFe] diagram.  The standard procedure is outlined in Graves \&
Schiavon (2008) and is applied to a sample of local Universe ellipticals using
stacked SDSS spectra in Graves \etal (2009).  Their results can be summarized is that
ellipticals (from $M_J = -21$ to $-25$) increase in $\alpha$/Fe with luminosity,
increase in age (from 6 to 12 Gyrs) and increase in [Fe/H] (from $-$0.4 to nearly
solar).  They also present a CMR for the SDSS color $g-r$ that progresses from 0.720
at $M_r = -18.8$ ($M_J = -20.7$) to 0.795 at $M_r = -23.0$ ($M_J = -25.1$).  Compared
to the CMR in Table 3, this agrees well at the faint end (0.721 for our sample), but
deviates significantly at the high luminosity end (0.829 for our sample).

Other Lick/IDS indices work find similar results but the number of young ellipticals
($\tau < 6$ Gyrs) varies from study to study.  For example, Kuntschner \etal (2010)
find very few ellipticals with ages less than 10 Gyrs.  Thomas \etal (2010) finds
10\% of their elliptical sample with ages around 3 Gyrs. Their Figure 7 displays
almost constant age for most of the sample (approximately 8 Gyrs) with 10\%
containing a much younger population.  These galaxies have $u-r < 2.4$, whereas only
6\% of our sample is that blue and all at the lowest luminosities in agreement with
the CMR.  McDermid \etal (2015) also finds a vast majority of ellipticals are old
($\tau > 10$ Gyrs).

Studies of high redshift ellipticals present a conflicting picture.  The massive end
of the galaxy mass spectrum grows considerably, presumably by mergers (Brammer \etal
2011), to become the local ellipticals observed by this study.  However, those
objects derive from a population beyond $z=2$ that displays a large range in color
and SFRs (van Dokkum \etal 2011).  Fumagalli \etal (2016) find that ellipticals at
$z=2$ have ages that correspond to local ages of 8 Gyrs, in agreement with the Graves
\etal ages.  And other Lick/IDS indices studies propose that age is correlated with
galaxy mass (Gallazzi \etal 2006), again in agreement with the observations of galaxy
evolution from high redshift.

With respect to metallicity, the Lick/IDS indices studies are more coherent.  They all
universally find increasing [Fe/H] with increasing luminosity, stellar mass and
velocity dispersion.  But the range of [Fe/H] values is significantly lower than
predicted by colors.  Most find [Fe/H] between $-$0.5 and solar for the range of
luminosities investigated in this paper, with the metallicity range decreasing with
later studies.  While some studies find $<$Fe$>$ or [MgFe] values well above solar
(e.g., Gallazzi \etal 2006), most find a majority of ellipticals with less than solar
metallicities, which would overlap their colors with globular clusters.

Most Lick/IDS indices studies are in consensus concerning the results from
$\alpha$/Fe observations such that they find a trend of increasing $\alpha$/Fe with
stellar mass.  While interpretation of the $\alpha$/Fe trends can be complicated, the
primary driver in $\alpha$/Fe is the duration of star formation.  Long durations lead
to low $\alpha$/Fe (due to the increasing contribution of Fe from SN Ia) and short
bursts lead to high $\alpha$/Fe values.  The trend of $\alpha$/Fe indicates, at the
very least, that the age difference between low and high mass ellipticals may arises
solely from a difference in the star formation duration (in the direction of younger
ages for low mass ellipticals).  The complication is that to achieve near solar
$\alpha$/Fe values only requires a duration of 1 to 2 Gyrs, a difference in age that
is difficult to discern in indices or colors.  Thomas \etal (2010) propose a scenario
where low mass ellipticals have a longer duration plus a rejuvenation phase several
Gyrs after the initial SF era, again, difficult to result in the red colors of even
low luminosity ellipticals. 

Lastly, many studies conclude that ellipticals host, to some degree, a young stellar
component (e.g. Trager \etal 2000; Kaviraj \etal 2007; Kuntschner \etal 2010).  The
amount of recent SF (and how recent) varies considerably and it is unclear whether
the Lick/IDS indices are more sensitive to recent SF than colors.  This would explain
the difference between index ages and color ages as they reflect differing effects
from a young component.  We do know that any SF event involving more than 5\% of the
stellar mass of an elliptical within the last 500 Myrs would alter the colors in a
fashion greater than the photometric errors quoted herein, particularly for the $NUV$
and $u$ filters.

With respect to stellar mass in ellipticals, a majority of the Lick/IDS indices
studies are in agreement that high mass ellipticals are all old (12 Gyrs) and high in
metallicity (although not as metal-rich as deduced from their colors).  Combined with
red colors in all the filters (except $NUV$), it is assumed that high mass
ellipticals formed from a monolithic scenario, an initial short of duration less than
0.5 Gyrs with a halt of SF by galactic winds.  This is the canonical scenario where
the range in [Fe/H] by galaxy mass is due solely to the depth of the gravitational
well and the longer time required before the galactic winds gain enough energy to
remove the remaining gas and halt SF.

The formation scenarios proposed for low mass ellipticals vary from study to study.
For example, McDermid \etal (2015) propose two phase elliptical star formation
history starting with an early monolithic collapse phase producing metal-rich,
$\alpha$/Fe enhanced ellipticals followed by an era accreting gas for an extended era
of star formation. This produces a low mass population of elliptical that is younger
and metal-poor (from the infalling gas) while the massive ellipticals quench early
(for older ages).  Chiosi \& Carraro (2002) also claim that line indices results can
be modeled by a scenario where high mass ellipticals form by monolithic processes and
low mass ellipticals are the result of irregular and intermittent episodes of star
formation.  Here the observed lower age indices are interpreted as the result of a
more complex star formation history in low mass ellipticals (perhaps a series of
bursts), particular longer durations of initial star formation, which would match the
observed $\alpha$/Fe ratios.  The difference in the evolution of low versus high mass
ellipticals is supported by the Faber \etal (2007) observations that found rapid
changes in the number of red, presumably, early-type galaxies since $z=1$.   While it
appears that most massive galaxies have already joined the red sequence at $z=1$
(Treu et al. 2005; Bundy et al. 2006; Cimatti et al. 2006), the low mass end of the
red sequence seems to be subject to the continued arrival and rejuvenation of
galaxies up to the present day (Treu \etal 2005; Schawinski \etal 2007).

The key conflict, outlined by this paper, is that the results from the Lick/IDS
line-strength studies can not be reconciled with the average colors of present-day
ellipticals from $NUV$ to 3.6$\mu$m.  And the claims that various formation and
evolution scenarios reproduces observations ignores the colors of ellipticals.  With
respect to ages, we find agreement with results from Lick/IDS indices on the high
mass end for old ($\tau = 12$ Gyrs) ages.  However, the conjecture of ages from 4 to
6 Gyrs for the low mass end of the elliptical sequence is in direct contradiction
with the observed colors.  While this comparison is model dependent (i.e., the
indices values and expected colors) it is important to note that both indices and
colors are extracted from the same SSP models.  And, while colors fade faster than
indices for young populations, the model predicted colors are still significantly
deviant from the indices extracted ages.  This is shown in \S3.3 for all the colors
in our sample.

This type of color analysis has only be lightly explored once the Lick/IDS
line-strength system was defined in the literature.  For example, Chiosi \& Carraro
(2002) consider the effect of secondary episodes of SF on $B-V$ colors and rule out
any activity with 5 Gyrs and engaging more than 5\% of the galaxy mass.  And Rakos
\etal (2008) consider the impact of younger ages in cluster ellipticals using narrow
band optical colors.  The latter study outlined the incompatible with ages younger
than 10 Gyrs and mean cluster elliptical colors.

In addition to the difficulty of reconciling mean colors with ages, the CMR across
our filter sets is also incompatible with even a very narrow range of ages outside 12
Gyrs.  One can reproduce the slope of each individual CMR by adjusting age downward
with a corresponding increase in metallicity.  But the resulting age and metallicity
values produce discrepant values in the other colors and their CMR's.  In fact, as
shown in \S5, all the CMR's across all colors are explained solely by changes in
[Fe/H] and any introduction of younger age on the low luminosity end forces the
slopes of other CMR's outside the photometric and fit errors.  While Kaviraj \etal
(2005) argues, through a $\Lambda$CDM hierarchical merger model, that the CMR is not
a meaningful tool for testing monolithic versus merger scenarios, we find, in fact,
that a comparison across a large range in wavelength does highly constrain the
possible range in age and metallicity.

The results of this study of the colors of ellipticals can be summarized as the
following:

\begin{itemize}

\item{} Multi-color photometry is presented for a large sample of local ellipticals
selected by morphology and isolation.  The sample uses data from $GALEX$, SDSS, 2MASS
and {\it Spitzer} to cover the filters $NUV$, $ugri$, $JHK$ and 3.6$\mu$m with
various levels of completeness outlined in Table 1.

\item{} Excellent agreement (typically less than 0.1 mags) is found between
magnitudes measured herein and the colors and magnitudes from the RC3 and other
published photometry.  Metric colors defined by the total luminosity in the $J$ band
are defined for all galaxies in the sample.  Mean colors for various luminosity
ranges are listed in Table 2.

\item{} The various two-color diagrams are very coherent from color to color, meaning
that galaxies defined to be red in one color are always red in other colors.  The
exception is the behavior of $NUV-u$, which is discussed separately in \S6.  The
colors of ellipticals are extensions of globular cluster colors with a detectable change
in slope of each two-color relationship from the globulars.

\item{} Various 12 Gyrs SSP models from the literature are tested against the
two-color diagrams.  All reproduce globular cluster colors with varying degrees of
accuracy with respect to metallicity.  Most importantly, the ellipticals have a
shallower slope not predicted by the SSP models.  The difference is greater for bluer
colors.  This is strong evidence that ellipticals are only roughly approximated
by an SSP model and a composite population, with a range of metallicities, is
required to explain the colors of ellipticals (see Rakos \& Schombert 2009).  

\item{} The simplest composite population is one of singular age and a range of
metallicities outlined by a chemical enrichment model.  We use a model with an age of
12 Gyrs and a chemical enrichment scenario outlined in Schombert \& McGaugh (2014).
This first order composite model is sufficient to match all the colors in our sample
except for the $NUV-u$ colors.  Most importantly, only a 12 Gyrs aged population with
varyingly metallicity is required to match ellipticals colors, and ages less than 8
Gyrs are ruled out based on colors unless extremely high metallicities are assumed.
In addition, a mean [Fe/H] value can be assigned based on the average of five colors
with the same level of accuracy as the globular calibration.

\item{} Using the globular clusters to calibrate the color-metallicity relationship,
we find the metallicities of ellipticals are expected to range from [Fe/H]=$-$0.5 to
+0.4, much higher than the [Fe/H] values found by Lick/IDS indices values.  And no
evidence for less than 10 Gyrs ages, particularly younger ages for low mass
ellipticals.  In addition, comparison of Lick/IDS indices and colors plus composite
population models are difficult to reconcile with young galaxy ages.  While it is not
the intent of this study to deduce age or metallicity values strictly from colors, we
do find that the deduced ages and metallicity from line studies are in conflict with
the expected colors.

\item{} The CMR across all colors display the increasing color with luminosity
(except $NUV-u$, see below).  The measured slopes agree with slopes from numerous
other studies on the CMR from the near-UV to the near-IR.  Modeling the CMR using
younger ages requires unrealistic metallicities at the low luminosity end and, more
importantly, it is impossible to reproduce the CMR across all colors with younger
age population even using metallicity as a variable.

\item{} The $NUV$ colors have unusual behavior near $u$ with a inverse CMR and
up-and-down signature in two-color diagrams.  This indicates an decrease in the UV
upturn to intermediate luminosity ellipticals, then a strengthening to higher masses.
Models with BHB tracks can reproduce this behavior and matching the globular cluster
colors indicating the UV upturn is a metallicity effect.

\end{itemize}

The tension between galaxy colors and Lick/IDS indices was first outlined in
Schombert \& Rakos (2009) and is even more salient with this study in terms of
improved accuracy and a wider wavelength coverage.  On the other hand, the measured
line indices values have also improved and there is no indication that the published
values are anything less than increasingly accurate and multiple observations
reinforce the reliability of those values.  Improved models do not resolve the
conflict and, as most of our conclusions about the formation and evolution of
galaxies are based on these types of observations.   It is critical that future
indices work include colors (even if extracted from the spectra) as a reality check
to the continuum measurements of the SED in galaxies.  A combination of colors and
indices are required to explore the effects of low luminosity components (e.g., a
weak metal-poor component or a recent SF event).

\noindent Acknowledgements: 

We thank the editorial staff and the referee, Guy Worthey, for their diligence and
insightful comments.  We are also grateful to Stacy McGaugh and Federico Lelli for
encouragement.  The software and funding for this project was supported by NASA's
Applied Information Systems Research (AISR) and Astrophysics Data Analysis Program
(ADAP) programs.  Data used for this study was based on observations made with (1)
the NASA Galaxy Evolution Explorer, GALEX is operated for NASA by the California
Institute of Technology under NASA contract NAS5-98034, (2) SDSS where funding has
been provided by the Alfred P.  Sloan Foundation, the Participating Institutions, the
National Science Foundation, the U.S. Department of Energy, the National Aeronautics
and Space Administration, the Japanese Monbukagakusho, the Max Planck Society, and
the Higher Education Funding Council for England, (3) the Two Micron All Sky Survey
(2MASS), which is a joint project of the University of Massachusetts and the Infrared
Processing and Analysis Center/California Institute of Technology, funded by the
National Aeronautics and Space Administration and the National Science Foundation and
(4) archival data obtained with the Spitzer Space Telescope, which is operated by the
Jet Propulsion Laboratory, California Institute of Technology under a contract with
NASA.  In addition, this research has made use of the NASA/IPAC Extragalactic
Database (NED) which is operated by the Jet Propulsion Laboratory, California
Institute of Technology, under contract with the National Aeronautics and Space
Administration. 

\pagebreak

\clearpage
\renewcommand{\arraystretch}{0.8}

\begin{deluxetable}{lccccccccc}
\tablecolumns{10}
\small
\tablewidth{0pt}
\tablecaption{Number of Galaxies per Filter Match}

\tablehead{
\\
\colhead{} &
\colhead{} &
\colhead{$u$} &
\colhead{$g$} &
\colhead{$r$} &
\colhead{$i$} &
\colhead{$J$} &
\colhead{$H$} &
\colhead{$K$} &
\colhead{3.6} \\

}

\startdata

{\it GALEX} $NUV$ & $\vert$  &  106  &  106  &  106  &  106  &  149  &  149  &  149  &  101 \\
SDSS $u$          & $\vert$  &       &  263  &  263  &  263  &  263  &  263  &  263  &   89 \\
SDSS $g$          & $\vert$  &       &       &  263  &  263  &  263  &  263  &  263  &   89 \\
SDSS $r$          & $\vert$  &       &       &       &  263  &  263  &  263  &  263  &   89 \\
SDSS $i$          & $\vert$  &       &       &       &       &  263  &  263  &  263  &   89 \\
2MASS $J$         & $\vert$  &       &       &       &       &       &  436  &  436  &  149 \\
2MASS $H$         & $\vert$  &       &       &       &       &       &       &  436  &  149 \\
2MASS $K$         & $\vert$  &       &       &       &       &       &       &       &  149 \\

\enddata
\end{deluxetable}

\begin{deluxetable}{cccccc}
\tablecolumns{6}
\small
\tablewidth{0pt}
\tablecaption{Mean Elliptical Colors and Metallicities}

\tablehead{
\\
\colhead{$L$} &
\colhead{2.0$L_*$} &
\colhead{1.0$L_*$} &
\colhead{0.5$L_*$} &
\colhead{0.2$L_*$} &
\colhead{0.1$L_*$} \\
}

\startdata

$M_J$  & $-$24.6  & $-$23.8  & $-$23.1  & $-$22.1  & $-$21.3  \\
& & & & & \\
$NUV-u$ & 3.027 & 3.137 & 3.207 & 3.171 & 3.211 \\
$\sigma$ & 0.183  & 0.169  & 0.140  & 0.181  & 0.169 \\
$N$ & 17  & 31  & 26  & 14  & 18 \\
& & & & & \\
$u-g$ & 1.850 & 1.834 & 1.804 & 1.767 & 1.741 \\
$\sigma$ & 0.061  & 0.049  & 0.059  & 0.053  & 0.101 \\
$N$ & 67  & 92  & 52  & 30  & 22 \\
& & & & & \\
$g-r$ & 0.829 & 0.816 & 0.801 & 0.775 & 0.741 \\
$\sigma$ & 0.024  & 0.021  & 0.030  & 0.028  & 0.047 \\
$N$ & 67  & 92  & 52  & 30  & 22 \\
& & & & & \\
$r-i$ & 0.424 & 0.415 & 0.408 & 0.401 & 0.378 \\
$\sigma$ & 0.017  & 0.015  & 0.014  & 0.009  & 0.022 \\
$N$ & 67  & 92  & 52  & 30  & 22 \\
& & & & & \\
$i-J$ & 1.623 & 1.609 & 1.594 & 1.562 & 1.535 \\
$\sigma$ & 0.046  & 0.031  & 0.049  & 0.032  & 0.068 \\
$N$ & 67  & 92  & 52  & 30  & 22 \\
& & & & & \\
$J-K$ & 0.981 & 0.965 & 0.950 & 0.923 & 0.912 \\
$\sigma$ & 0.031  & 0.030  & 0.036  & 0.031  & 0.036 \\
$N$ & 106  & 155  & 94  & 48  & 30 \\
& & & & & \\
$K-3.6$ & 0.168 & 0.154 & 0.148 & 0.132 & 0.129 \\
$\sigma$ & 0.032  & 0.019  & 0.030  & 0.020  & 0.026 \\
$N$ & 30  & 36  & 36  & 23  & 24 \\
& & & & & \\

$<$Fe/H$>$ & 0.30 & 0.26 & 0.20 & 0.13 & $-$0.02 \\
           & $\pm0.05$ & $\pm0.06$ & $\pm0.04$ & $\pm0.04$ & $\pm0.05$ \\

\enddata

\end{deluxetable}

\begin{deluxetable}{ccccc}
\tablecolumns{5}
\small
\tablewidth{0pt}
\tablecaption{CMR Properties}

\tablehead{
\\
\colhead{color} &
\colhead{$\Delta$(color/mag)} &
\colhead{$\sigma$} &
\colhead{$<$Fe/H$>$} &
\colhead{$<$Fe/H$>$} 
\\
&
&
&
\colhead{@ $M_J = -26$} &
\colhead{@ $M_J = -20$} 
\\

}

\startdata

$NUV-u$ & $+$0.0247 & 0.17 & $-$ & $-$ \\
$u-g$   & $-$0.0387 & 0.06 & +0.38 & $-$0.16 \\
$g-r$   & $-$0.0247 & 0.03 & +0.39 & $-$0.20 \\
$r-i$   & $-$0.0116 & 0.02 & +0.37 & $-$0.07 \\
$i-J$   & $-$0.0333 & 0.05 & $-$  & $-$    \\
$J-K$   & $-$0.0181 & 0.04 & +0.32 & $-$0.08 \\
$K-3.6$ & $-$0.0119 & 0.04 & $-$  & $-$    \\
$g-K$   & $-$0.0878 & 0.04 & +0.36 & $-$0.10 \\

\enddata
\end{deluxetable}

\clearpage

\begin{deluxetable}{rlccccc}
\tablecolumns{7}
\small
\tablewidth{0pt}
\tablecaption{Models of Young Elliptical Ages}

\tablehead{
\\
\colhead{Age} &
\colhead{Fe/H]} &
\colhead{$u-g$} &
\colhead{$g-r$} &
\colhead{$r-i$} &
\colhead{$g-K$} &
\colhead{$J-K$}
\\

}

\startdata

$M_J$ = & $-$21.3 & +1.705 & +0.735 & +0.378 & +3.510 & +0.901 \\
\\

 6 Gyrs & $-$0.02 & +1.535 & +0.650 & +0.336 & +3.250 & +0.860 \\
\hline
 $\Delta$ &     & $-$0.170 & $-$0.085 & $-$0.042 & $-$0.260 & $-$0.041 \\
\\
 6 Gyrs & +0.30   & +1.620 & +0.735 & +0.377 & +3.594 & +0.943 \\
\hline
  $\Delta$ &     & $-$0.085 & +0.000 & $-$0.001 & +0.084 & +0.042 \\

\\
$M_J$ = & $-$23.1 & +1.775 & +0.780 & +0.399 & +3.668 & +0.933 \\
\\

 9 Gyrs & +0.0 & +1.624 & +0.695 & +0.356 & +3.388 & +0.881 \\
\hline
  $\Delta$ &   & $-$0.151 & $-$0.085 & $-$0.043 & $-$0.280 & $-$0.052 \\
\\
 9 Gyrs & +0.31 & +1.729 & +0.779 & +0.398 & +3.707 & +0.954 \\
\hline
  $\Delta$ &   & $-$0.046 & $-$0.001 & $-$0.001 & +0.039 & +0.021 \\
\\

\enddata
\end{deluxetable}

\end{document}